\documentclass[prd,10pt,aps]{revtex4-2}
\usepackage{graphicx,epstopdf}
\usepackage[caption=false]{subfig}
\usepackage{appendix}
\usepackage[T1]{fontenc}
\usepackage{lmodern}
\usepackage[dvipsnames,x11names]{xcolor}
\usepackage[colorlinks=true,linkcolor=Magenta!80!black,citecolor=Magenta!70!black,urlcolor=Magenta!80!black]{hyperref}
\usepackage[sort&compress]{natbib}
\usepackage{lipsum}
\usepackage{morefloats}
\usepackage[pdf]{pstricks}
\usepackage{amsmath}
\usepackage{amssymb}
\usepackage{amsfonts}
\usepackage{rotating}
\usepackage{cancel}
\usepackage{mathtools}
\usepackage{bbm}
\usepackage{dsfont}
\usepackage{bbold}
\usepackage{multirow}
\usepackage{ulem}
\usepackage{physics}
\usepackage{orcidlink}
\usepackage{colortbl}

\def\pslash{p\!\!\!\slash }
\def\qslash{q\!\!\!\slash }
\def\xslash{x\!\!\!\slash }
\def\eslash{\varepsilon\!\!\!\slash }

\hypersetup{
    pdfencoding=auto,
    psdextra
}

\begin{document}

\title{Analytic electromagnetic signatures of compact pentaquark structure: \\ A multi-current QCD light-cone sum rules analysis of the $P_{\psi s}^{\Lambda}$ states}

\author{Ula\c{s} \"{O}zdem\orcidlink{0000-0002-1907-2894}}%
\email[]{ulasozdem@aydin.edu.tr}
\affiliation{Health Services Vocational School of Higher Education, Istanbul Aydin University, Sefakoy-Kucukcekmece, 34295 Istanbul, T\"{u}rkiye}

\begin{abstract}
Probing the internal organization of hidden-charm pentaquarks---including the spin--color correlations that distinguish compact diquark--diquark--antiquark configurations from loosely bound hadronic molecules---requires observables that go beyond mass spectroscopy. We argue that multi-current QCD light-cone sum rules provide a diagnostic framework for this problem, not only through the expected sensitivity of the magnetic moment to the internal spin structure but, more importantly, through exact analytic relations among the flavor-sector contributions that are enforced by the algebra of the interpolating currents themselves. We identify two such signatures that emerge from our analysis of compact diquark--diquark--antiquark configurations: (i) for any of the four currents considered the light-quark contributions satisfy the exact ratio $\mu_{u}/\mu_{d} = e_{u}/e_{d} = -2$, reflecting a common Lorentz--color kernel that couples $u$ and $d$ only through their electric charges; and (ii) for the $J_{3}(x)$ current the charm-quark contribution vanishes identically, $\mu_{c}=0$, arising not from the pseudoscalar embedding of the charm diquark alone---since the same embedding occurs in $J_{1}(x)$, which yields a nonzero $\mu_{c}$---but from the specific Dirac structure associated with the anti-charm coupling in the global current, which produces an analytic cancellation in the OPE. We illustrate these signatures through explicit calculations using four independent diquark--diquark--antiquark interpolating currents $J_{1}(x)$--$J_{4}(x)$, assumed to carry $J^{P}=\tfrac{1}{2}^{-}$, and obtain $\mu_{J_{1}}=-1.35^{+0.35}_{-0.28}\,\mu_{N}$, $\mu_{J_{2}}=3.14^{+0.65}_{-0.50}\,\mu_{N}$, $\mu_{J_{3}}=1.01^{+0.25}_{-0.20}\,\mu_{N}$, and $\mu_{J_{4}}=-1.79^{+0.41}_{-0.34}\,\mu_{N}$. These four predictions are then assigned to the observed $P_{\psi s}^{\Lambda}(4338)$ and $P_{\psi s}^{\Lambda}(4459)$ resonances on the basis of the mass predictions of recent QCD sum rule analyses; we emphasize, however, that the corresponding $\pm 0.11~\text{GeV}$ mass uncertainties accommodate either state within $1\sigma$ of all four currents, so this current-to-state mapping is adopted only as an auxiliary phenomenological choice. The predicted magnitudes $|\mu|\sim 1$--$3\,\mu_{N}$ lie systematically above quark-model and heavy pentaquark chiral perturbation theory expectations ($|\mu|\lesssim 0.5\,\mu_{N}$). Applying the same flavor-decomposition procedure to two previous molecular LCSR analyses of the same states yields $\mu_{u}/\mu_{d}=-1/2$ rather than $-2$, providing an LCSR-internal contrast that operates at the flavor-decomposed level even when total magnitudes are comparable. Crucially, the two analytic signatures identified here are immune to the ambiguity of the state-to-current pairing and therefore offer falsifiable tests of the compact picture that survive the unavoidable phenomenological assumptions of mass-based assignments.
  \end{abstract}

\maketitle

\section{Introduction}\label{motivation}

Exotic hadronic configurations, such as pentaquarks and tetraquarks, have emerged as central topics in particle physics since the early formulation of the quark model~\cite{Gell-Mann:1964ewy}. Although neither the quark model nor quantum chromodynamics (QCD) forbids their existence, these states remained hypothetical for decades and were the subject of extensive theoretical and experimental studies. A major breakthrough occurred in 2003, when the Belle Collaboration reported the first observation of a candidate tetraquark state, the $X(3872)$, thereby marking the beginning of a new era in the exploration of exotic hadrons~\cite{Belle:2003nnu}.  The discovery of exotic hadrons composed of five valence quarks, known as pentaquarks, was first reported by the LHCb Collaboration in 2015. Two states, $P_{\psi}^N(4380)$ and $P_{\psi}^N(4450)$, were observed in the $J/\psi p$ decay channel~\cite{LHCb:2015yax}. Subsequent studies in 2019, based on a larger dataset, revealed that the $P_{\psi}^N(4450)$ resonance is actually two separate states, $P_{\psi}^N(4440)$ and $P_{\psi}^N(4457)$, and also identified a new resonance, $P_{\psi}^N(4312)$~\cite{LHCb:2019kea}. The existence of $P_{\psi}^N(4380)$ remains uncertain, as it has neither been confirmed nor definitively ruled out. In 2020, the LHCb Collaboration reported a hidden-charm pentaquark candidate, $P_{\psi s}^{\Lambda}(4459)$, in the $J/\psi \Lambda$ invariant mass spectrum via the decay $\Xi_b^0 \to J/\psi \Lambda K^-$~\cite{LHCb:2020jpq}, which was subsequently supported by evidence from the Belle Collaboration~\cite{Belle:2025pey}.  In 2021, the LHCb Collaboration reported the detection of a new pentaquark state in the $J/\psi p$ invariant mass spectrum~\cite{LHCb:2021chn}, whose resonance parameters differ from those of the previously established pentaquarks observed in the $\Lambda_b \rightarrow J/\psi p K$ decay channel. In 2022, the LHCb Collaboration reported the observation of a new pentaquark candidate, $P_{\psi s}^{\Lambda}(4338)$, in the $J/\psi \Lambda$ invariant mass spectrum by analyzing the decay process $B^- \to J/\psi \Lambda \bar{p}$~\cite{LHCb:2022ogu}. Concerning spin-parity assignments, an amplitude analysis by LHCb favors $J^P = \tfrac{1}{2}^-$ for the $P_{\psi s}^{\Lambda}(4338)$, whereas for the $P_{\psi s}^{\Lambda}(4459)$ the current data do not allow a unique determination and both $\tfrac{1}{2}^-$ and $\tfrac{3}{2}^-$ remain compatible with the measurements. Throughout this work we adopt $J^P = \tfrac{1}{2}^-$ as a working hypothesis for both states, with the understanding that our results for the $P_{\psi s}^{\Lambda}(4459)$ are conditional on this assignment. 

Despite these experimental advances, the internal structure of the $P_{\psi}^N$ and $P_{\psi s}^{\Lambda}$ states remains an open question. Several scenarios have been proposed, including a compact pentaquark composed of a diquark--diquark--antiquark configuration, or alternatively, a loosely bound hadronic molecular state formed by the interaction of a charmed baryon and a meson (for reviews, see \cite{Liu:2024uxn,Meng:2022ozq,Chen:2022asf}). A crucial feature of the current landscape is that the measured masses alone do not discriminate among these competing structural scenarios: compact diquark--diquark--antiquark configurations, hadronic molecules, and quark-model bound states all yield mass predictions consistent with the observed $4338$ and $4459$~MeV values within their respective uncertainties, largely because these states lie near common hadronic thresholds that constrain the total energy more strongly than any particular internal configuration. The situation is further complicated by the persistent ambiguity in the spin-parity assignment of the $P_{\psi s}^{\Lambda}(4459)$, where both $J^P = \tfrac{1}{2}^-$ and $\tfrac{3}{2}^-$ remain compatible with existing data. Under these conditions, observables that go beyond the mass spectrum---and that respond differently to distinct internal configurations even when the corresponding mass predictions coincide---become indispensable for any meaningful structural discrimination. A detailed review of experimental findings and theoretical developments for both confirmed and candidate pentaquark states can be found in~\cite{Esposito:2014rxa, Esposito:2016noz, Olsen:2017bmm, Lebed:2016hpi, Nielsen:2009uh,  Brambilla:2019esw, Agaev:2020zad, Chen:2016qju, Ali:2017jda, Guo:2017jvc, Liu:2019zoy, Yang:2020atz, Dong:2021juy, Dong:2021bvy, Meng:2022ozq, Chen:2022asf}.

Electromagnetic multipole moments are a natural candidate for such complementary observables: existing theoretical predictions for the magnetic dipole moment of the $P_{\psi s}^{\Lambda}$ states span a wide range across competing structural scenarios, with quark-model and heavy pentaquark chiral perturbation theory analyses typically yielding $|\mu|\lesssim 0.5\,\mu_{N}$ and QCD light-cone sum rules-based studies producing magnitudes that range from a few tenths to a few $\mu_{N}$ depending on the assumed internal organization of the state. We will see that the comparison among these scenarios is more subtle than a simple magnitude-based separation would suggest, and that a coherent structural diagnosis requires going beyond the total magnetic moment to its flavor-decomposed contributions. They are directly sensitive to the spatial arrangement of quarks and to their spin orientations within the hadron, and they respond differently to compact and extended internal configurations. For hidden-charm pentaquarks in particular, the magnetic dipole moment provides information about the alignment of quark spins and their interactions with external magnetic fields, while higher multipoles probe finer anisotropies of the charge and current densities. Existing studies of electromagnetic multipole moments for pentaquarks remain comparatively limited~\cite{Ozdem:2025jda,  Ozdem:2024rch,  Ozdem:2024rqx,  Ozdem:2023htj,  Ozdem:2022kei, Wang:2016dzu,  Ortiz-Pacheco:2018ccl, Xu:2020flp, Ozdem:2018qeh,  Ozdem:2021ugy, Li:2021ryu, Gao:2021hmv, Guo:2023fih, Wang:2022nqs, Wang:2022tib, Ozdem:2024jty, Li:2024wxr, Li:2024jlq,  Mutuk:2024ltc, Mutuk:2024jxf, Mutuk:2024ach, Ozdem:2024usw, Ozdem:2025fks, Zhu:2025abk, Ozdem:2026gmn, Ozdem:2025ion,Mutuk:2026zxp}, and the results reported span a wide range, reflecting the strong dependence of these observables on the assumed internal configuration.

Motivated by these considerations, the present work employs the multi-current QCD light-cone sum rules (LCSR) framework~\cite{Chernyak:1990ag, Braun:1988qv, Balitsky:1989ry} to compute the magnetic moments of the $P_{\psi s}^{\Lambda}(4338)$ and $P_{\psi s}^{\Lambda}(4459)$ pentaquarks, under the assumption of a compact diquark--diquark--antiquark configuration. We construct four independent interpolating currents $J_{1}(x)$--$J_{4}(x)$ that encode different spin--flavor organizations of the constituent diquarks, and extract both the total magnetic moments and their decomposition into individual quark-sector contributions. The use of multiple interpolating currents serves a dual purpose: it quantifies the sensitivity of the magnetic moment to the internal diquark structure---an aspect that has been repeatedly emphasized in LCSR studies of exotic hadrons---and, as we shall see in Sec.~\ref{numerical}, the resulting flavor decomposition exhibits structural regularities whose origin lies in the algebra of the interpolating currents and which can be traced analytically. These features, together with the explicit numerical values, form the main content of our results. Although directly measuring the magnetic moment of a short-lived hadronic state such as the $P_{\psi s}^{\Lambda}$ states poses significant experimental challenges, it remains a feasible objective in future high-luminosity facilities. One promising approach involves analyzing the photoproduction cross-section in electron--proton collisions: the magnetic moment, together with other electromagnetic form factors, governs the coupling of the pentaquark to the electromagnetic field, and consequently the production rate of the $P_{\psi s}^{\Lambda}$ via photon-induced reactions is sensitive to this coupling. Complementary information can be extracted from the angular distributions of the decay products in photoproduction events, since spin-dependent interactions influenced by the magnetic moment leave characteristic imprints on these distributions.

The article is organized as follows: Section \ref{formalism} presents the sum rule derivation for the magnetic moments of the $P_{\psi s}^{\Lambda}$ states. Section \ref{numerical} presents the numerical results for the magnetic moments and their flavor-sector decomposition, followed by a concise summary of the main findings in Section \ref{summary}.

\section{Sum rule derivation} \label{formalism}

The analysis of the electromagnetic structure of the $P_{\psi s}^{\Lambda}$ states, within the formalism of LCSR, is initiated through the formulation of the following correlation functions:
\begin{align}
 \label{edmn00}
 \Pi _{ \alpha }(p,q)&=-\int d^{4}x \,e^{ip\cdot x}\int d^{4}y\,e^{iq \cdot y}\,
\langle 0|\mathcal{T}\Big\{J(x) J_{\alpha}^\gamma(y)
\bar J(0)\Big\}|0\rangle.
\end{align}%
In this expression, $ J_\alpha^\gamma (y) $ represents the electromagnetic current, while $ J(x) $ denotes the interpolating current corresponding to the $P_{\psi s}^{\Lambda}$ states, which carries the quantum numbers $ J^P = \frac{1}{2}^- $.

The correlation function may alternatively be expressed as
\begin{align} \label{edmn01}
\Pi(p,q)=i\int d^4x \, e^{ip \cdot x} \langle 0 | T\{ J(x) \bar{J}(0) \} | 0 \rangle_F \,.
\end{align}

It should be emphasized that $\Pi(p,q)$ represents $\varepsilon^\alpha \Pi_\alpha(p,q)$. In this context, $ F $ denotes the external background electromagnetic field (EBGEM), with 
\begin{equation}
F_{\alpha\beta} = i (\varepsilon_\alpha q_\beta - \varepsilon_\beta q_\alpha) e^{-iq \cdot x},
\end{equation}
where $ \varepsilon_\beta $ and $ q_\alpha $ correspond to the polarization and four-momentum of the photon, respectively. 

The EBGEM approach offers a notable advantage by enabling a gauge-invariant separation of soft and hard photon contributions~\cite{Ball:2002ps}. Moreover, the EBGEM is treated as an infinitesimally weak background field, allowing the correlation function in Eq.~(\ref{edmn01}) to be systematically expanded in powers of the field strength, leading to the following representation:
\begin{align}
\Pi (p,q) &= \Pi^{(0)}(p,q) + \Pi^{(1)}(p,q)+\cdots  .
\end{align}
In this decomposition, $\Pi^{(0)}(p,q)$ denotes the correlation function in the absence of an external background electromagnetic field, which is associated with the mass sum rules and lies beyond the scope of the present analysis. In contrast, $\Pi^{(1)}(p,q)$ accounts for contributions arising from single-photon emission~\cite{Ball:2002ps,Novikov:1983gd,Ioffe:1983ju}. Accordingly, within the framework of LCSR, the extraction of the electromagnetic multipole moments of the relevant hadrons requires only the computation of the $\Pi^{(1)}(p,q)$ component.

With these clarifications established, we now proceed to derive the LCSR for the magnetic moments of the $P_{\psi s}^{\Lambda}$ states. The initial step of our analysis involves computing the hadronic representation of the correlation function. Within the hadronic framework, by inserting complete sets of $P_{\psi s}^{\Lambda}$ states carrying the same quantum numbers as the interpolating currents, we obtain
 \begin{align}\label{edmn02}
\Pi^{Had}(p,q)&=\frac{\langle0\mid J(x) \mid
{\mathrm{P_{\psi s}^{\Lambda}}}(p, s) \rangle}{[p^{2}-m_{\mathrm{P_{\psi s}^{\Lambda}}}^{2}]}
\langle {\mathrm{P_{\psi s}^{\Lambda}}}(p, s)\mid
{\mathrm{P_{\psi s}^{\Lambda}}}(p+q, s)\rangle_F 
\frac{\langle {\mathrm{P_{\psi s}^{\Lambda}}}(p+q, s)\mid
\bar J(0) \mid 0\rangle}{[(p+q)^{2}-m_{\mathrm{P_{\psi s}^{\Lambda}}}^{2}]}+ \cdots . 
\end{align}

For the subsequent calculations, the matrix elements in Eq.~(\ref{edmn02}) are required. These can be expressed in terms of hadronic quantities, such as the spinor $ u(p,s) $ and the residue $ \lambda_{\mathrm{P_{\psi s}^{\Lambda}}} $, as illustrated below:
\begin{align}
\langle0\mid J(x)\mid {\mathrm{P_{\psi s}^{\Lambda}}}(p, s)\rangle=&\lambda_{\mathrm{P_{\psi s}^{\Lambda}}} \gamma_5 \, u(p,s),\label{edmn04}\\
\langle {\mathrm{P_{\psi s}^{\Lambda}}}(p+q, s)\mid\bar J(0)\mid 0\rangle=&\lambda_{\mathrm{P_{\psi s}^{\Lambda}}} \gamma_5 \, \bar u(p+q,s)\label{edmn004}
.
\end{align}

For a spin-$\frac{1}{2}$ hadron, the electromagnetic current matrix element can be represented using Lorentz-invariant form factors ($ F_i(q^2) $). The corresponding general expression is provided in~\cite{Leinweber:1990dv}: 
\begin{align}
\langle {\mathrm{P_{\psi s}^{\Lambda}}}(p, s)\mid {\mathrm{P_{\psi s}^{\Lambda}}}(p+q, s)\rangle_F &=\varepsilon^\mu\,\bar u(p, s)\,\Gamma_\mu \,u(p+q, s), \label{edmn005}
\end{align}
 where 
 \begin{align}
\Gamma_\mu = \big[F_1(q^2)
+F_2(q^2)\big] \gamma_\mu +F_2(q^2)
\frac{(2p+q)_\mu}{2 m_{\mathrm{P_{\psi s}^{\Lambda}}}}.
 \end{align}

By combining the above expressions and performing a summation over spins, one obtains the resulting equations for the correlation functions in the hadronic representation:
\begin{align}
\label{edmn05}
\Pi^{Had}(p,q)&= \frac{\lambda^2_{\mathrm{P_{\psi s}^{\Lambda}}}}{[p^{2}-m_{{\mathrm{P_{\psi s}^{\Lambda}}}}^{2}][(p+q)^{2}-m_{{\mathrm{P_{\psi s}^{\Lambda}}}}^{2}]} \bigg[\Big(F_1(q^2)+F_2(q^2)\Big)\Big(
  2 (\varepsilon . p) \pslash -
  m_{\mathrm{P_{\psi s}^{\Lambda}}}\,\eslash \pslash
  -m_{\mathrm{P_{\psi s}^{\Lambda}}}\,\eslash \qslash
  +\pslash\eslash\qslash
  \Big)
  \bigg]. 
\end{align}

In order to determine the magnetic form factor, $ F_M(q^2) $, of the $P_{\psi s}^{\Lambda}$ state, it is essential to relate it to the form factors $ F_i(q^2) $. The relevant expressions are given below:
\begin{align}
\label{edmn07}
F_M(q^2) &= F_1(q^2) + F_2(q^2).
\end{align} 
Using the expressions described above, the electromagnetic form factors of the $P_{\psi s}^{\Lambda}$ states can be determined. In the case of a real photon, where $ q^2 = 0 $, the relevant form factor can be directly related to the magnetic moment. The corresponding relation is given below:
\begin{align}
\label{edmn08}
\mu_{\mathrm{P_{\psi s}^{\Lambda}}} &= \frac{ e}{2\, m_{\mathrm{P_{\psi s}^{\Lambda}}}} \,F_M(0).
\end{align}
With this relation, the hadronic representation of the correlation function is complete. We are now prepared to proceed to the QCD-side evaluation of the correlation function.

 One of the most essential ingredients in the QCD representation of the correlation function is the construction of an interpolating current that can effectively couple to the hadronic states under consideration.  In constructing suitable interpolating currents, it is essential to align the quark field configurations with the valence content and quantum numbers of hidden-charm pentaquark states. The attractiveness of the one-gluon exchange mechanism naturally favors the formation of diquarks in the color antitriplet channel. Previous QCD sum-rule analyses \cite{Wang:2010sh, Kleiv:2013dta} indicate that scalar and axial-vector diquark correlations are the most favorable configurations. Guided by these considerations, we employ in this study interpolating currents of the axial-vector--diquark--scalar--diquark--antiquark type and the axial-vector--diquark--axial-vector--diquark--antiquark type. 
 Taking into account the assumption that the pentaquark studied in this work has the quantum numbers $J^P = \frac{1}{2}^-$, the possible interpolating currents can be written as follows: 
\begin{align}\label{curpcs2}
J_{1}(x)&= \varepsilon^{abc}\varepsilon^{ade} \varepsilon^{bfg}\Big\{ \big[ {u}^T_d(x) C \gamma_5 {d}_e(x) \big] \big[ {s}^T_f(x) C \gamma_5 c_g(x)\big]   \Big\}  C  \bar{c}^{T}_{c}(x) \, , \\
J_{2}(x)&= \varepsilon^{abc}\varepsilon^{ade} \varepsilon^{bfg}\Big\{ \big[ {u}^T_d(x) C \gamma_5 {d}_e(x) \big] \big[ {s}^T_f(x) C \gamma_\mu c_g(x)\big]   \Big\}   \gamma_5 \gamma^\mu C  \bar{c}^{T}_{c}(x) \, , \\
J_{3}(x)&=\frac{\varepsilon^{abc}\varepsilon^{ade} \varepsilon^{bfg}}{\sqrt{2}} \Big\{ \big[ {u}^T_d(x) C \gamma_\mu {s}_e(x) \big] \big[ {d}^T_f(x) C \gamma_5 c_g(x)\big]   - 
\big[ {d}^T_d(x) C \gamma_\mu {s}_e(x) \big] \big[ {u}^T_f(x) C \gamma_5c_g(x)\big]  \Big\}  \gamma_5 \gamma^\mu C  \bar{c}^{T}_{c}(x) \, , \\
J_{4}(x)&=\frac{\varepsilon^{abc}\varepsilon^{ade} \varepsilon^{bfg}}{\sqrt{2}} \Big\{ \big[ {u}^T_d(x) C \gamma_\mu {s}_e(x) \big] \big[ {d}^T_f(x) C \gamma^\mu c_g(x)\big]   - 
\big[ {d}^T_d(x) C \gamma_\mu {s}_e(x) \big] \big[ {u}^T_f(x) C \gamma^\mu c_g(x)\big]  \Big\}  C  \bar{c}^{T}_{c}(x) \, , 
\end{align}
where $a$, $b$, $\cdots$ are color indexes and the $C$ is the charge conjugation operator.

All four interpolating currents considered here are, in principle, capable of coupling to the relevant pentaquark states. In~\cite{Wang:2025fqh}, the corresponding hadron masses were evaluated using these four currents: $m^{J_{1}} = 4.47 \pm 0.11~\text{GeV}$ and $m^{J_{2}} = 4.51 \pm 0.10~\text{GeV}$ for the scalar-diquark--containing currents $J_{1}(x)$ and $J_{2}(x)$, and $m^{J_{3}} = 4.33 \pm 0.11~\text{GeV}$ and $m^{J_{4}} = 4.37 \pm 0.11~\text{GeV}$ for the currents $J_{3}(x)$ and $J_{4}(x)$ built on a vector light--strange diquark. On the basis of these central values, the lower-mass pair $\{J_{3}(x),J_{4}(x)\}$ is phenomenologically paired with the $P_{\psi s}^{\Lambda}(4338)$, and the higher-mass pair $\{J_{1}(x),J_{2}(x)\}$ with the $P_{\psi s}^{\Lambda}(4459)$. We emphasize, however, that this assignment is far from unique: the $\pm 0.11~\text{GeV}$ mass uncertainties comfortably accommodate either state within the $1\sigma$ range of all four currents, and a cross-assignment (or even a mixing of currents coupling to a single physical state) cannot be excluded on mass grounds alone. We therefore adopt this pairing as a working scheme rather than as a definitive structural identification. As we shall see in Sec.~\ref{numerical}, the structural features that emerge from the flavor decomposition of the magnetic moments are properties of the interpolating currents themselves rather than of any particular state-to-current assignment, and would survive a reshuffling of this mapping; the mass-based pairing adopted here therefore enters only as an auxiliary phenomenological choice. Accordingly, we present calculations for the $P_{\psi s}^{\Lambda}(4338)$ magnetic moment using the $J_{3}(x)$ and $J_{4}(x)$ currents, and for the $P_{\psi s}^{\Lambda}(4459)$ using the $J_{1}(x)$ and $J_{2}(x)$ currents.

The QCD side evaluation of the correlation function is carried out in the deep Euclidean region, where $ p^2 \ll 0 $ and $ (p+q)^2 \ll 0 $. Within this kinematic regime, the correlation function can be expressed in terms of photon distribution amplitudes (DAs). To obtain this representation, the interpolating currents are substituted into the correlation function defined in Eq.~(\ref{edmn00}), and applying Wick's theorem allows the correlation function to be recast in terms of both heavy and light quark propagators as follows:
\begin{align} \label{QCD1}
\Pi^{QCD-J_1(x)}(p,q)&= -i\,\varepsilon^{abc}\varepsilon^{a^{\prime}b^{\prime}c^{\prime}}\varepsilon^{ade} \varepsilon^{a^{\prime}d^{\prime}e^{\prime}}\varepsilon^{bfg} \varepsilon^{b^{\prime}f^{\prime}g^{\prime}}
\int d^4x\, e^{ip\cdot x}\,
\langle 0|
\Big\{
  \mbox{Tr}\Big[  \gamma_5 S_{d}^{ee^\prime}(x) \gamma_5 \widetilde S_{u}^{dd^\prime }(x) \Big] \nonumber\\
&\times
 \mbox{Tr}\Big[ \gamma_\mu S_c^{gg^\prime}(x) \gamma_\nu \widetilde S_{s}^{ff^\prime }(x) \Big]  
\Big \} 
\Big( \gamma_5 \gamma^\mu   \widetilde S_c^{c^{\prime}c } (-x)  \gamma^\nu   \gamma_5 \Big)
|0 \rangle_F , \\[6pt]
%
\Pi^{QCD-J_2(x)}(p,q)&= -i\,\varepsilon^{abc}\varepsilon^{a^{\prime}b^{\prime}c^{\prime}}\varepsilon^{ade} \varepsilon^{a^{\prime}d^{\prime}e^{\prime}}\varepsilon^{bfg} \varepsilon^{b^{\prime}f^{\prime}g^{\prime}}
\int d^4x\, e^{ip\cdot x}\,
\langle 0|
\Big\{
  \mbox{Tr}\Big[  \gamma_5 S_{d}^{ee^\prime}(x) \gamma_5 \widetilde S_{u}^{dd^\prime }(x) \Big] \nonumber\\
&\times
 \mbox{Tr}\Big[ \gamma_5 S_c^{gg^\prime}(x) \gamma_5 \widetilde  S_{s}^{ff^\prime }(x) \Big]  
 \Big \} 
\Big(  \widetilde S_c^{c^{\prime}c } (-x)  \Big)
|0 \rangle_F ,  \label{QCD2}\\[6pt]
%
\Pi^{QCD-J_3(x)}(p,q)&= -\frac{i}{2}\,\varepsilon^{abc}\varepsilon^{a^{\prime}b^{\prime}c^{\prime}}\varepsilon^{ade} \varepsilon^{a^{\prime}d^{\prime}e^{\prime}}\varepsilon^{bfg} \varepsilon^{b^{\prime}f^{\prime}g^{\prime}}
\int d^4x\, e^{ip\cdot x}\,
\langle 0|
\Big\{
  \mbox{Tr}\Big[  \gamma_\mu S_{s}^{ee^\prime}(x) \gamma_\nu  \widetilde S_{u}^{dd^\prime }(x) \Big] \nonumber\\
&\times
 \mbox{Tr}\Big[ \gamma_5 S_c^{gg^\prime}(x) \gamma_5   \widetilde S_{d}^{ff^\prime }(x) \Big] 
 + \mbox{Tr}\Big[  \gamma_\mu S_{s}^{ee^\prime}(x) \gamma_\nu  \widetilde S_{d}^{dd^\prime }(x) \Big]
 \mbox{Tr}\Big[ \gamma_5 S_c^{gg^\prime}(x) \gamma_5 \widetilde S_{u}^{ff^\prime }(x)  \Big] 
\nonumber\\
&
 + \mbox{Tr} \Big[ \gamma_\mu S_s^{ee^\prime}(x) 
\gamma_\nu \widetilde S_{d}^{fd^\prime }(x)   \gamma_5 S_{c}^{gg^\prime}(x) \gamma_5 \widetilde  S_{u}^{df^\prime }(x) \Big]
\nonumber\\
&
 + \mbox{Tr} \Big[ \gamma_\mu S_s^{ee^\prime}(x) 
\gamma_\nu \widetilde S_{u}^{fd^\prime }(x)   \gamma_5 S_{c}^{gg^\prime}(x) \gamma_5 \widetilde S_{d}^{df^\prime }(x) \Big]
\Big \} 
\Big( \gamma_5 \gamma^\mu \widetilde S_c^{c^{\prime}c } (-x) \gamma^\nu   \gamma_5 \Big)
|0 \rangle_F , \label{QCD3} \\[6pt]
%
\Pi^{QCD-J_4(x)}(p,q)&= -\frac{i}{2}\,\varepsilon^{abc}\varepsilon^{a^{\prime}b^{\prime}c^{\prime}}\varepsilon^{ade} \varepsilon^{a^{\prime}d^{\prime}e^{\prime}}\varepsilon^{bfg} \varepsilon^{b^{\prime}f^{\prime}g^{\prime}}
\int d^4x\, e^{ip\cdot x}\,
\langle 0|
\Big\{
  \mbox{Tr}\Big[  \gamma_\mu S_{s}^{ee^\prime}(x) \gamma_\nu \widetilde  S_{u}^{dd^\prime }(x) \Big] \nonumber\\
&\times
 \mbox{Tr}\Big[ \gamma^\mu S_c^{gg^\prime}(x) \gamma^\nu \widetilde  S_{d}^{ff^\prime }(x) \Big] 
+ \mbox{Tr}\Big[  \gamma_\mu S_{s}^{ee^\prime}(x) \gamma_\nu \widetilde  S_{d}^{dd^\prime }(x) \Big]
 \mbox{Tr}\Big[ \gamma^\mu S_c^{gg^\prime}(x) \gamma^\nu  \widetilde S_{u}^{ff^\prime }(x)  \Big] 
\nonumber\\
&
 + \mbox{Tr} \Big[ \gamma_\mu S_s^{ee^\prime}(x) 
\gamma_\nu \widetilde S_{d}^{fd^\prime }(x)   \gamma^\mu S_{c}^{gg^\prime}(x) \gamma^\nu \widetilde  S_{u}^{df^\prime }(x) \Big]
\nonumber\\
&
 + \mbox{Tr} \Big[ \gamma_\mu S_s^{ee^\prime}(x) 
\gamma_\nu \widetilde S_{u}^{fd^\prime }(x) \gamma^\mu S_{c}^{gg^\prime}(x) \gamma^\nu \widetilde S_{d}^{df^\prime }(x) \Big]
\Big \} 
\Big( \widetilde S_c^{c^{\prime}c } (-x)  \Big)
|0 \rangle_F, \label{QCD4}
\end{align}
where $\widetilde S^{a a^{\prime} }_{q(c)} = C S^{a a^{\prime} \mathrm{T}}_{q(c)} C$. 

In these expressions, the light and charm quark propagators, denoted by $ S_q(x) $ and $ S_c(x) $, respectively, are given by the following forms~\cite{Balitsky:1987bk, Belyaev:1985wza}:
\begin{align}
\label{edmn13}
S_{q}(x)&= \frac{1}{2 \pi x^2}\Big(i \frac{\xslash}{x^2}- \frac{m_q}{2}\Big) 
-i\frac { g_s }{16 \pi^2 x^2} \int_0^1 du \, G^{\mu \nu} (ux)
\bigg[\bar u \rlap/{x} 
\sigma_{\mu \nu} + u \sigma_{\mu \nu} \rlap/{x}
 \bigg],\\
S_{c}(x)&=\frac{m_{c}^{2}}{4 \pi^{2}} \bigg[ \frac{K_{1}\big(m_{c}\sqrt{-x^{2}}\big) }{\sqrt{-x^{2}}}
+i\frac{{\xslash}~K_{2}\big( m_{c}\sqrt{-x^{2}}\big)}
{(\sqrt{-x^{2}})^{2}}\bigg]
-i\frac{m_{c}\,g_{s} }{16\pi ^{2}}  \int_0^1 du \,G^{\mu \nu}(ux)\bigg[ (\sigma _{\mu \nu }{\xslash}
+{\xslash}\sigma _{\mu \nu }) 
    \frac{K_{1}\big( m_{c}\sqrt{-x^{2}}\big) }{\sqrt{-x^{2}}}
   \nonumber\\
  &
 +2\sigma_{\mu \nu }K_{0}\big( m_{c}\sqrt{-x^{2}}\big)\bigg].
 \label{edmn14}
\end{align}%
In this context, $ G^{\mu\nu} (x) $ represents the background gluonic field strength tensor, while $ K_0\big( m_c \sqrt{-x^2} \big) $, $ K_1\big( m_c \sqrt{-x^2} \big) $, and $ K_2\big( m_c \sqrt{-x^2} \big) $ denote the modified Bessel functions of the second kind.

The correlation function receives contributions from two distinct sources: perturbative and non-perturbative. The perturbative component emerges when the photon couples to quarks via short-distance, perturbatively calculable processes, whereas the non-perturbative component stems from long-distance interactions between the photon and quarks. Calculating both contributions is crucial to obtain a complete and reliable description.

In evaluating the perturbative contributions, the propagator of one of the quarks is replaced by its expression that incorporates the photon interaction at the perturbative level.
\begin{align}
\label{free}
S^{free}(x) \rightarrow \int d^4y\, S^{free} (x-y)\,\rlap/{\!A}(y)\, S^{free} (y)\,, 
\end{align}
where $S^{free}(x)$ is the first term of the heavy and light quark propagators. 
The remaining propagators in Eqs.~(\ref{QCD1})–(\ref{QCD4}) are treated as free propagators. 

To incorporate non-perturbative effects, one of the light quark propagators in Eqs.~(\ref{QCD1})–(\ref{QCD4}) is replaced by the expression presented below:
 \begin{align}
\label{edmn21}
S_{\alpha\beta}^{ab}(x) \rightarrow -\frac{1}{4} \Big[\bar{q}^a(x) \Gamma_i q^b(0)\Big]\Big(\Gamma_i\Big)_{\alpha\beta},
\end{align}
where $\Gamma_i = \{\mathbb{1}, \gamma_5, \gamma_\mu, i\gamma_5 \gamma_\mu, \sigma_{\mu\nu}/2\}$. 
 Within this framework, the matrix elements $\langle \gamma(q) | \bar{q}(x) \Gamma_i G_{\alpha\beta} q(0) | 0 \rangle$ and $\langle \gamma(q) | \bar{q}(x) \Gamma_i q(0) | 0 \rangle$ arise, which are parameterized in terms of photon distribution amplitudes (DAs)~\cite{Ball:2002ps}. Photon DAs play a central role in incorporating non-perturbative QCD effects into the evaluation of correlation functions. In this work, we employ the explicit expressions given in \cite{Ball:2002ps}, which include contributions up to twist-4 accuracy. 
It is important to emphasize that the photon DAs considered here account exclusively for light-quark contributions. Although, in principle, long-distance photon emissions from charm quarks can occur, their effects are negligible in the present analysis. In particular, charm quark condensates are suppressed by the inverse heavy-quark mass, scaling as $\sim 1/m_c$~\cite{Antonov:2012ud}, and therefore provide a negligible contribution to the correlation function. Accordingly, in our analysis, we omit long-distance photon emissions from charm quark DAs and retain only the short-distance contributions, as specified in Eq.~(\ref{free}). By following the technical procedures outlined above, the QCD side expression for the magnetic moment is derived.

The magnetic moment of the $P_{\psi s}^{\Lambda}$ state is extracted within the LCSR framework by equating the correlation function expressed in terms of QCD parameters with its hadronic representation, employing the concept of quark--hadron duality. To efficiently suppress contributions from the continuum and higher resonances while enhancing the ground-state signal, continuum subtraction and Borel transformation are performed according to the standard LCSR procedure. The resulting magnetic moments, obtained through the complete set of steps outlined above, are summarized as follows:
\begin{align}
\label{edmn15}
\mu^{J_1(x)}_{P_{\psi s}^{\Lambda}} \,\lambda^{2, J_1(x)}_{P_{\psi s}^{\Lambda}} &=e^{\frac{m^2_{P_{\psi s}^{\Lambda}}}{\rm{M^2}}}\, \rho_1(\rm{M^2},\rm{s_0}),  ~~~~~~ 
\mu^{J_2(x)}_{P_{\psi s}^{\Lambda}} \,\lambda^{2, J_2(x)}_{P_{\psi s}^{\Lambda}} =e^{\frac{m^2_{P_{\psi s}^{\Lambda}}}{\rm{M^2}}}\, \rho_2 (\rm{M^2},\rm{s_0}),\\
\mu_{P_{\psi s}^{\Lambda}}^{J_3(x)} \,\lambda^{2, J_3(x)}_{P_{\psi s}^{\Lambda}} &=e^{\frac{m^2_{P_{\psi s}^{\Lambda}}}{\rm{M^2}}}\, \rho_3(\rm{M^2},\rm{s_0}) ~~~~~~~~
\mu_{P_{\psi s}^{\Lambda}}^{J_4(x)} \,\lambda^{2, J_4(x)}_{P_{\psi s}^{\Lambda}} =e^{\frac{m^2_{P_{\psi s}^{\Lambda}}}{\rm{M^2}}}\, \rho_4 (\rm{M^2},\rm{s_0}). \label{sondenk}
\end{align}

The explicit form of the functions $\rho_1 (\mathrm{M^2},\mathrm{s_0})$, $\rho_2 (\mathrm{M^2},\mathrm{s_0})$, $\rho_3 (\mathrm{M^2},\mathrm{s_0})$ and $\rho_4 (\mathrm{M^2},\mathrm{s_0})$ are presented in the appendix \ref{appa}.

\section{Results and discussions}\label{numerical}

The numerical analysis proceeds in two stages. We first fix the auxiliary parameters of the sum rule (the Borel mass parameter $\rm{M^2}$ and the continuum threshold $\rm{s_0}$) following the standard stability and convergence criteria, and extract the four magnetic moments associated with the interpolating currents $J_1(x)$--$J_4(x)$. We then turn to the central part of our analysis: the identification and discussion of two analytic signatures that emerge from the flavor decomposition of these results and that, as argued in the Introduction, constitute the principal outcome of the present work. The current-by-current physical interpretation and the comparison with existing theoretical approaches follow.

\subsection{Auxiliary parameters and numerical values}

The numerical values of the input parameters used in these computations 
are provided in Table~\ref{inputparameter}. The explicit definitions of 
the photon DAs ($\varphi_\gamma$, $\psi^v$, $\psi^a$, $\mathcal{A}$, 
$\mathcal{V}$, $h_\gamma$, $\mathbb{A}$, $\mathcal{S}$, 
$\tilde{\mathcal{S}}$, $\mathcal{T}_{1-4}$), the corresponding 
non-local matrix elements, and the numerical values of the additional 
DA parameters  entering these expressions are 
collected in Appendix~\ref{appb}, following the conventions of~\cite{Ball:2002ps}.
 \begin{table}[htb!]
	\addtolength{\tabcolsep}{10pt}
	\caption{List of input parameters utilized in our numerical computations.}
	\label{inputparameter}
\begin{tabular}{l|ccc}
               \hline\hline
               \\
Inputs & Values \\
\\
                                        \hline\hline
$m_s$&$ 93.5 \pm 0.8$~MeV \cite{ParticleDataGroup:2024cfk}   
                                               \\
$m_c$&$ 1.273 \pm 0.0046$~GeV \cite{ParticleDataGroup:2024cfk}
                                               \\
$m_{P_{\psi s}^{\Lambda}}^{J_1(x)}$&$ 4.47 \pm 0.11$~GeV \cite{Wang:2025fqh} 
                                \\
$m_{P_{\psi s}^{\Lambda}}^{J_2(x)}$&$ 4.51 \pm 0.10$~GeV \cite{Wang:2025fqh}
                                               \\
$m_{P_{\psi s}^{\Lambda}}^{J_3(x)}$&$ 4.33 \pm 0.11$~GeV \cite{Wang:2025fqh} 
                                \\
$m_{P_{\psi s}^{\Lambda}}^{J_4(x)}$&$ 4.37 \pm 0.11$~GeV \cite{Wang:2025fqh} 
                                \\
$\lambda_{P_{\psi s}^{\Lambda}}^{J_1(x)}  $&$ (1.86 \pm 0.30)\times 10^{-3} $~GeV$^6$ \cite{Wang:2025fqh}
                                \\
$ \lambda_{P_{\psi s}^{\Lambda}}^{J_2(x)}   $&$(3.43 \pm 0.55)\times 10^{-3} $~GeV$^6$ \cite{Wang:2025fqh}
                                \\
$\lambda_{P_{\psi s}^{\Lambda}}^{J_3(x)}  $&$ (2.81 \pm 0.47)\times 10^{-3} $~GeV$^6$ \cite{Wang:2025fqh}
                                \\
$ \lambda_{P_{\psi s}^{\Lambda}}^{J_4(x)}   $&$(2.34 \pm 0.42)\times 10^{-3} $~GeV$^6$ \cite{Wang:2025fqh}
                                \\
$\langle \bar qq\rangle $&$ (-0.24 \pm 0.01)^3 $~GeV$^3$ \cite{Ioffe:2005ym}
                                \\
$\langle \bar ss\rangle $&$ 0.8 \times (-0.24 \pm 0.01)^3 $~GeV$^3$ \cite{Ioffe:2005ym}
                                \\
$ \langle g_s^2G^2\rangle  $&$ 0.48 \pm 0.14 $~GeV$^4$ \cite{Narison:2018nbv}
                                \\
$f_{3\gamma} $&$ -0.0039 $~GeV$^2$ \cite{Ball:2002ps}
                                \\
$\chi $&$ -2.85 \pm 0.5 $~GeV$^{-2}$ \cite{Rohrwild:2007yt}
                                \\
                                      \hline\hline
 \end{tabular}
\end{table}

\begin{figure}[htb!]
\includegraphics[width=0.32\textwidth]{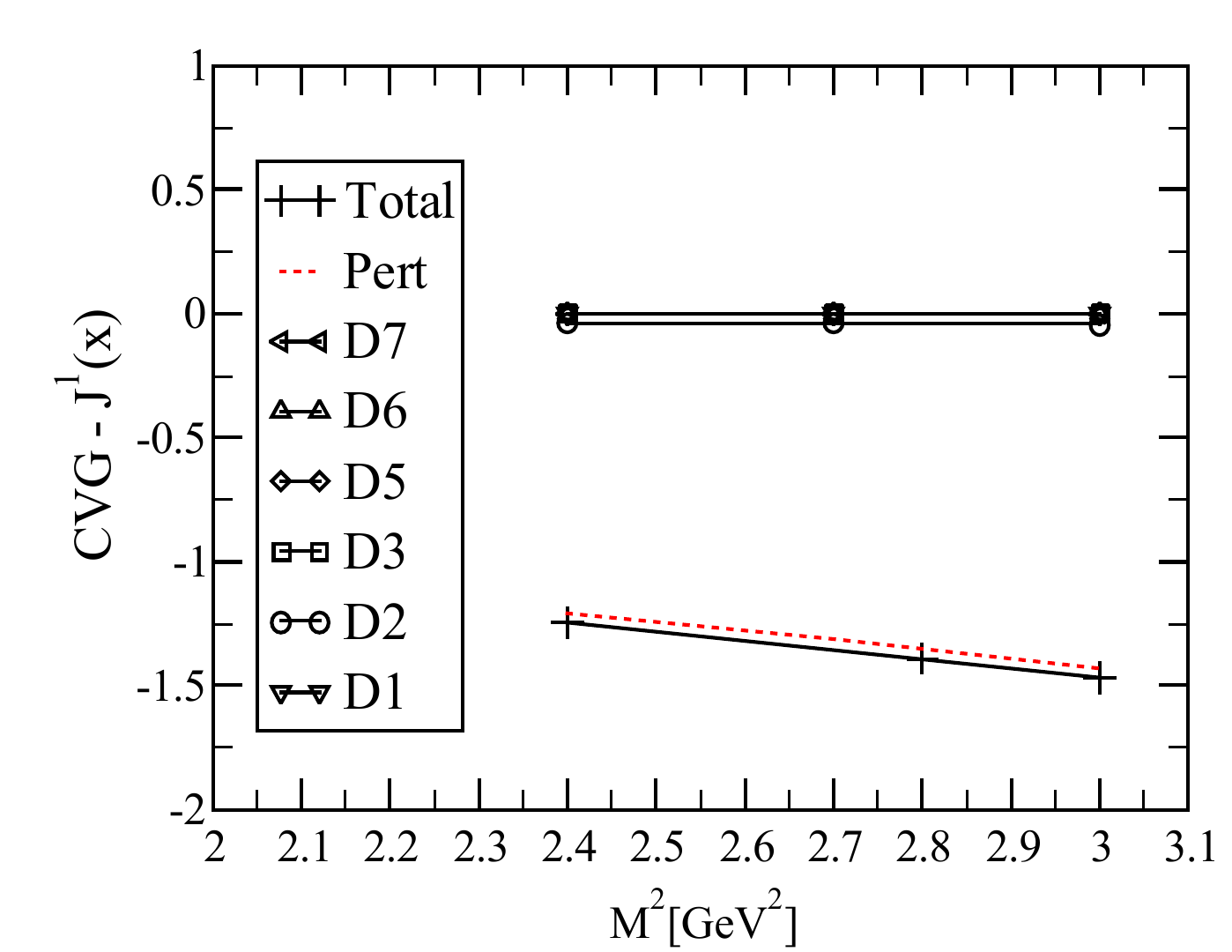}~~
\includegraphics[width=0.32\textwidth]{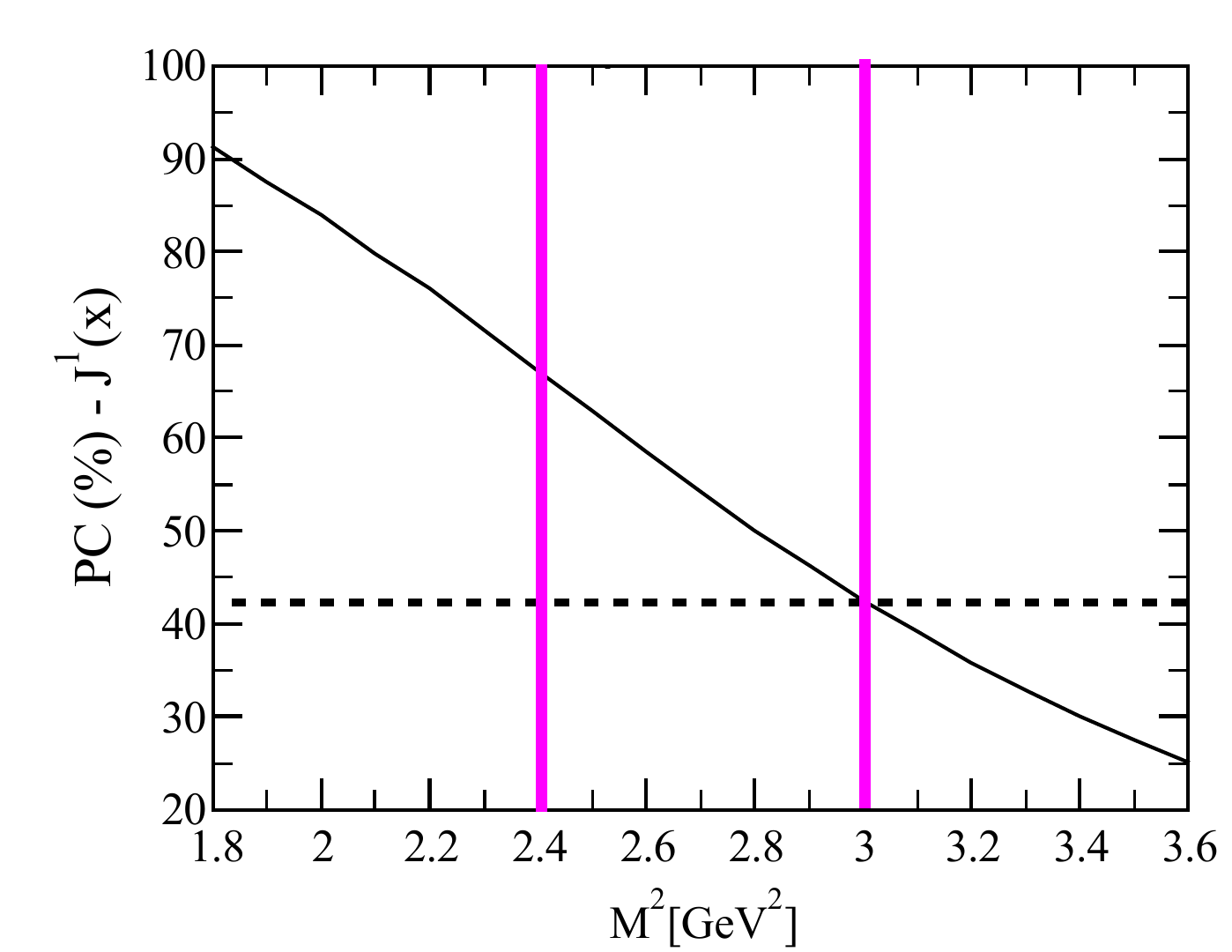}~~
\includegraphics[width=0.32\textwidth]{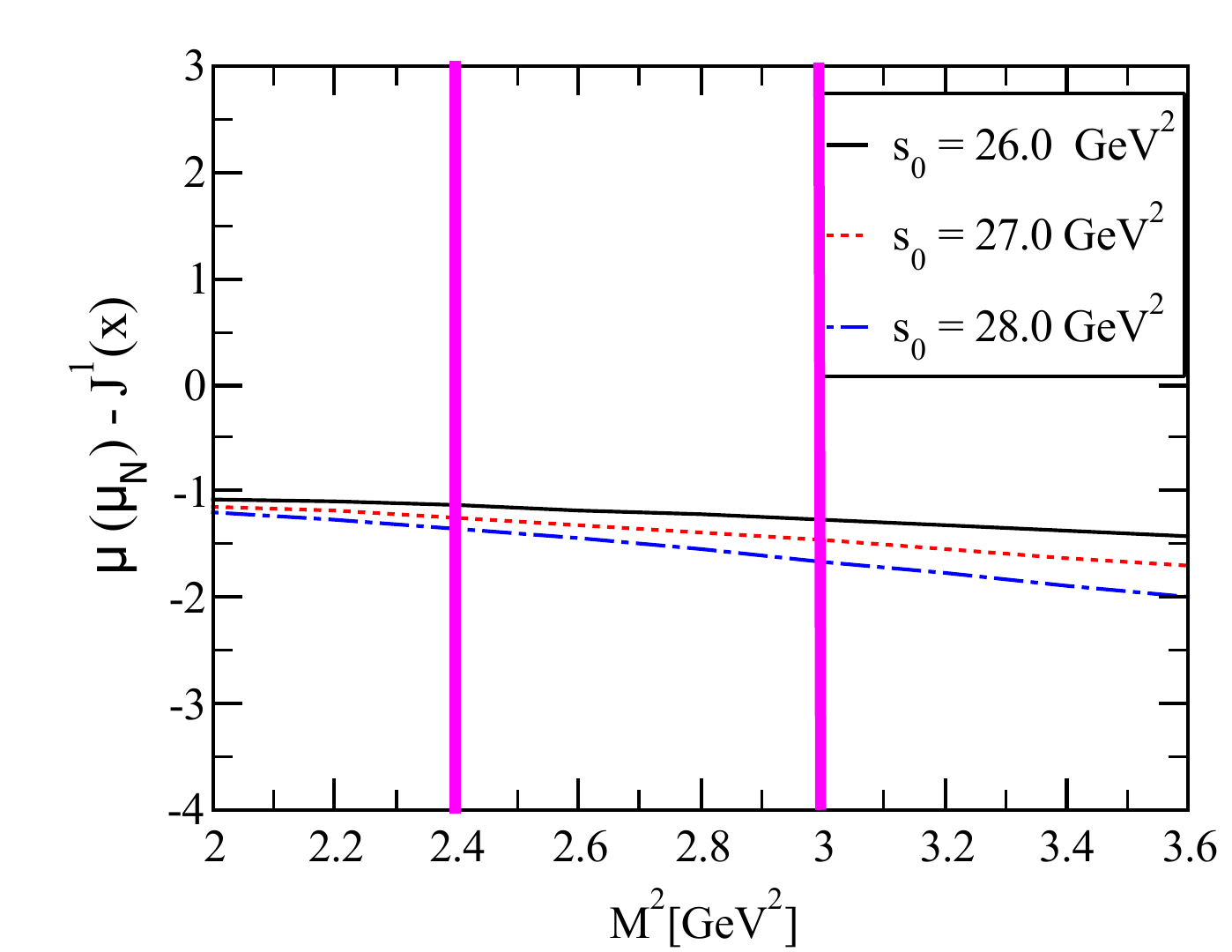}\\
\vspace{0.5 cm}
\includegraphics[width=0.32\textwidth]{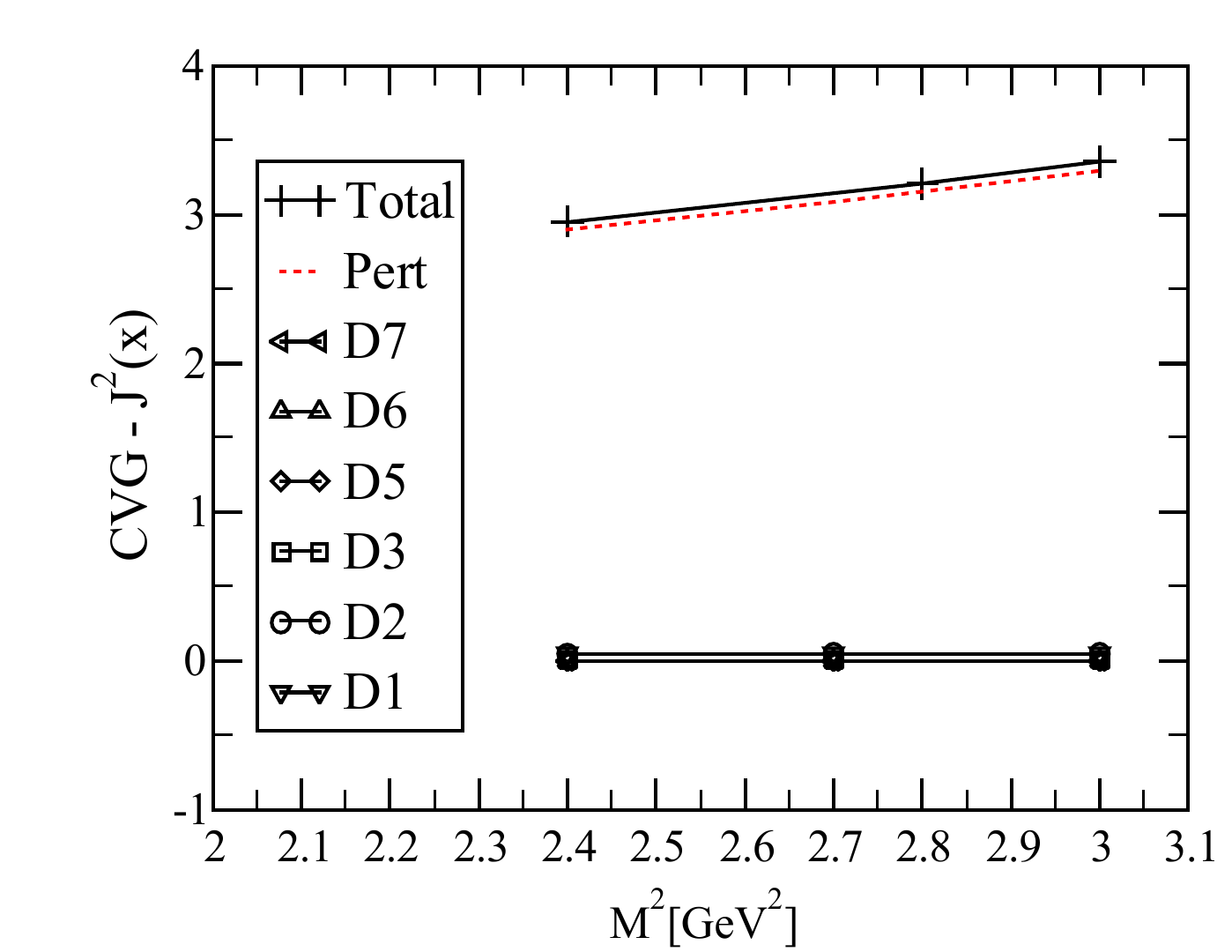}~~
\includegraphics[width=0.32\textwidth]{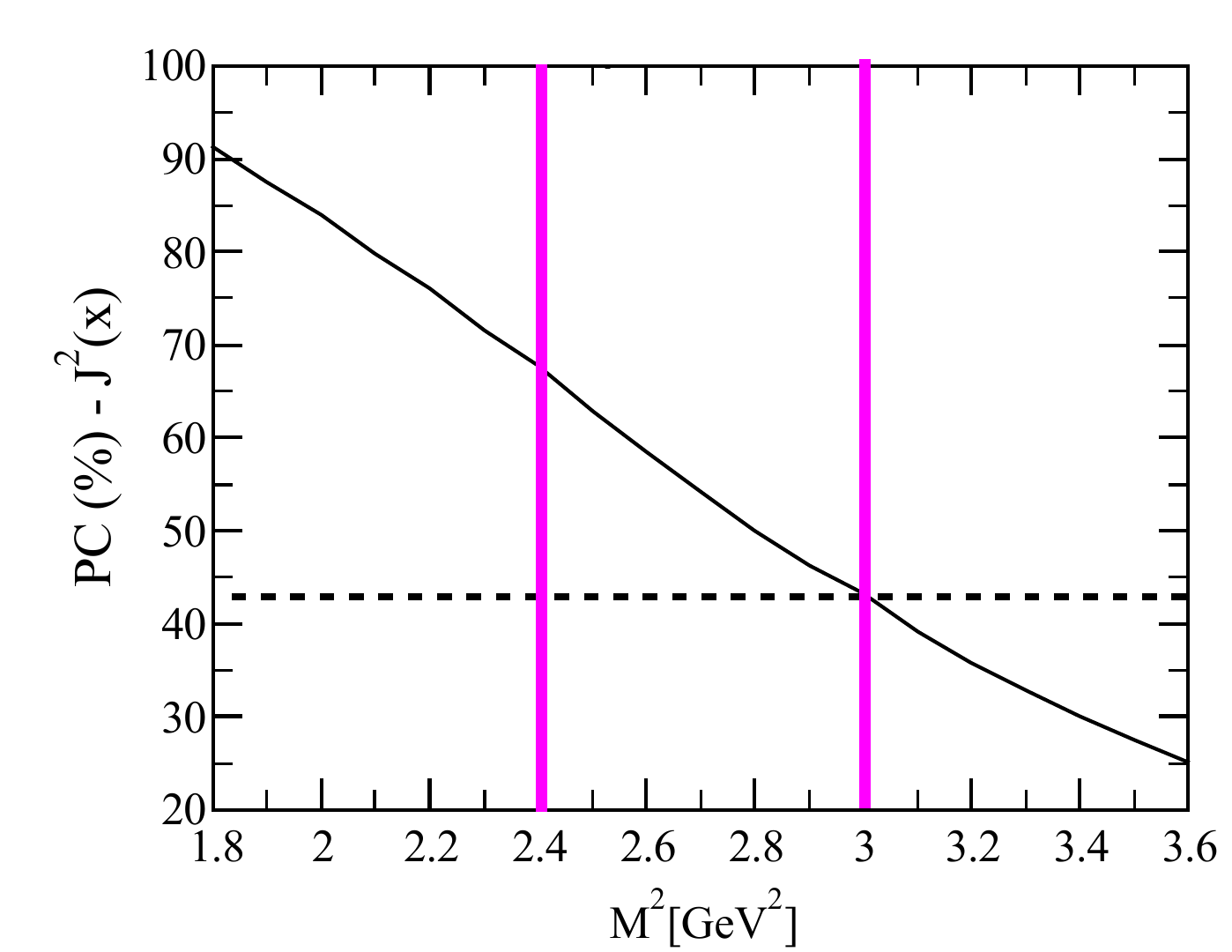}~~
\includegraphics[width=0.32\textwidth]{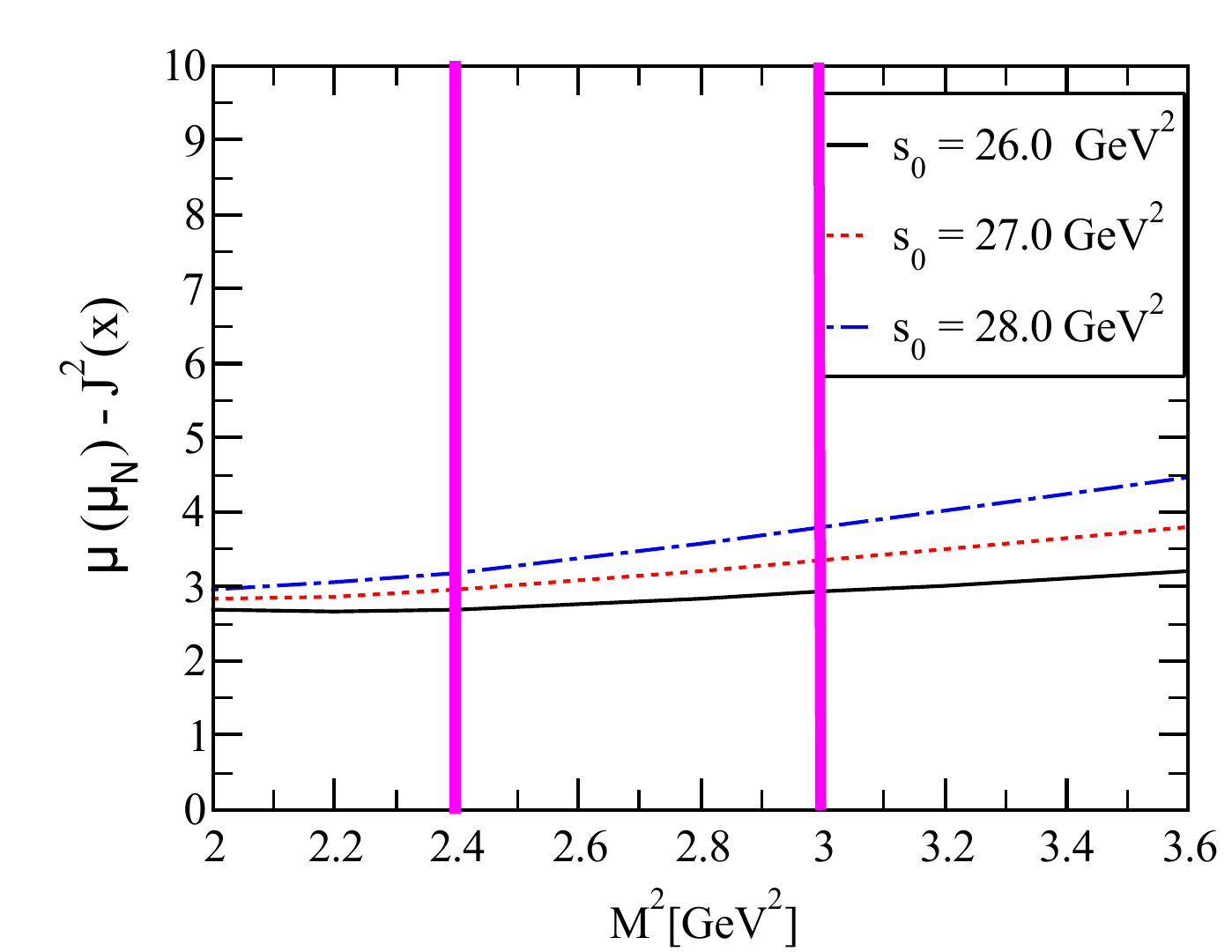}\\
\vspace{0.5 cm}
\includegraphics[width=0.32\textwidth]{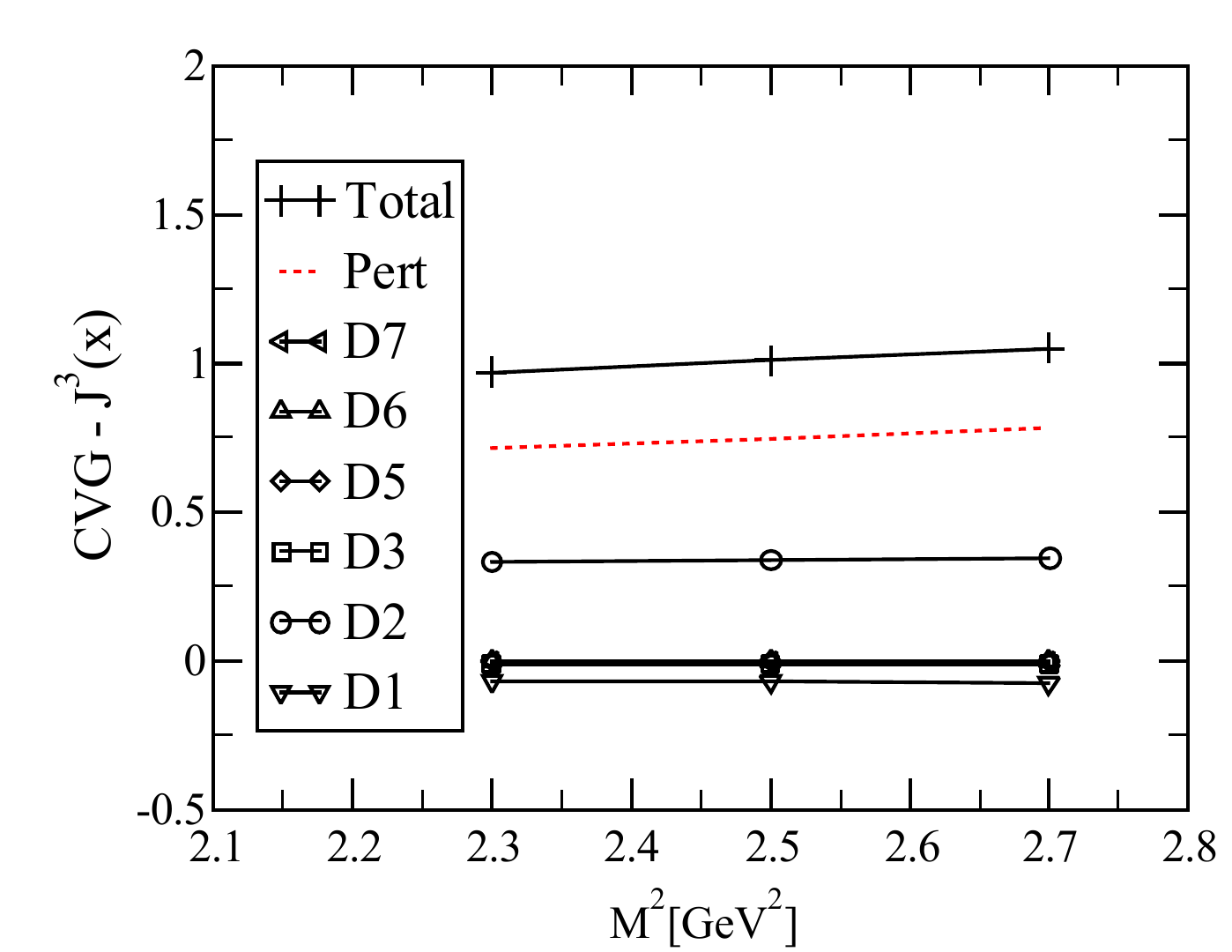}~~
\includegraphics[width=0.32\textwidth]{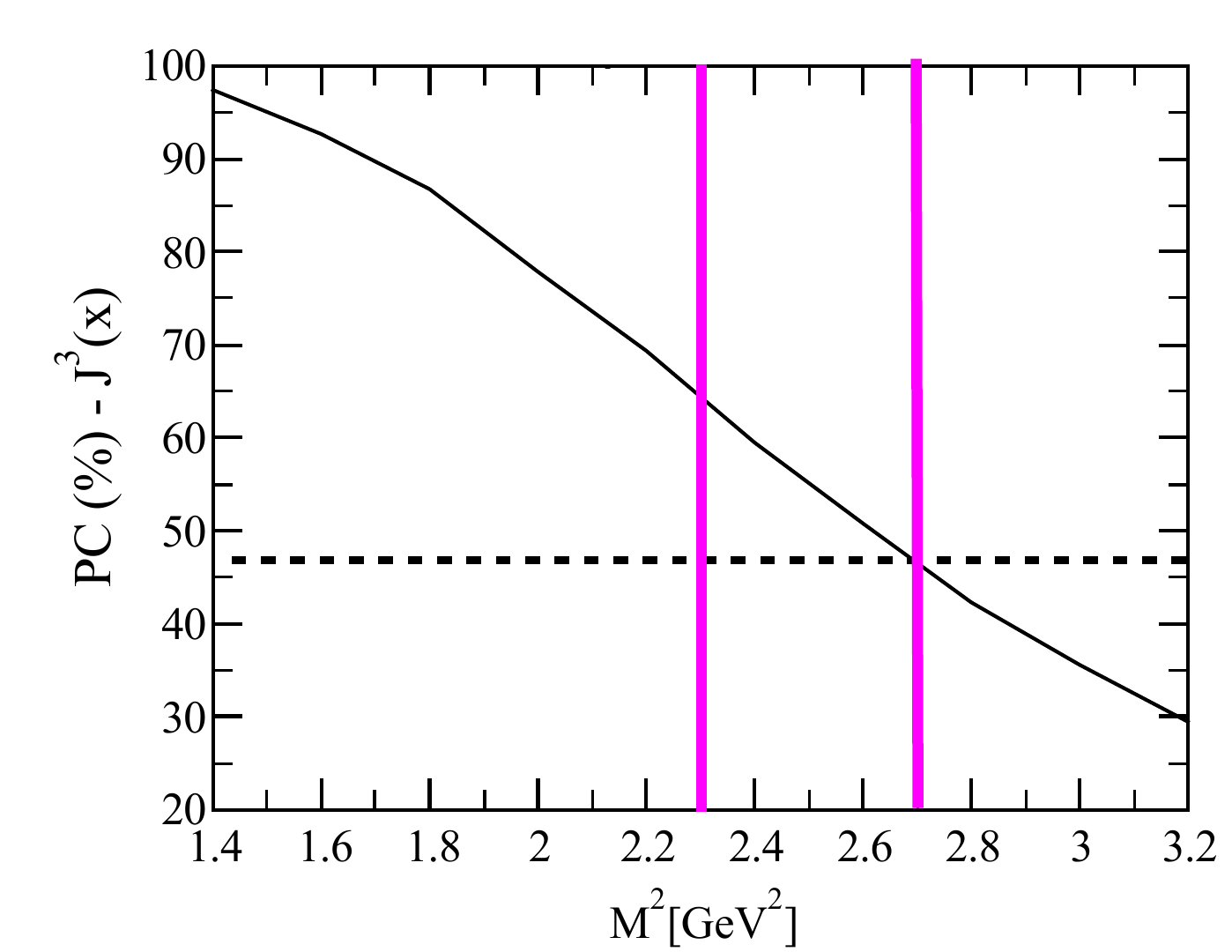}~~
\includegraphics[width=0.32\textwidth]{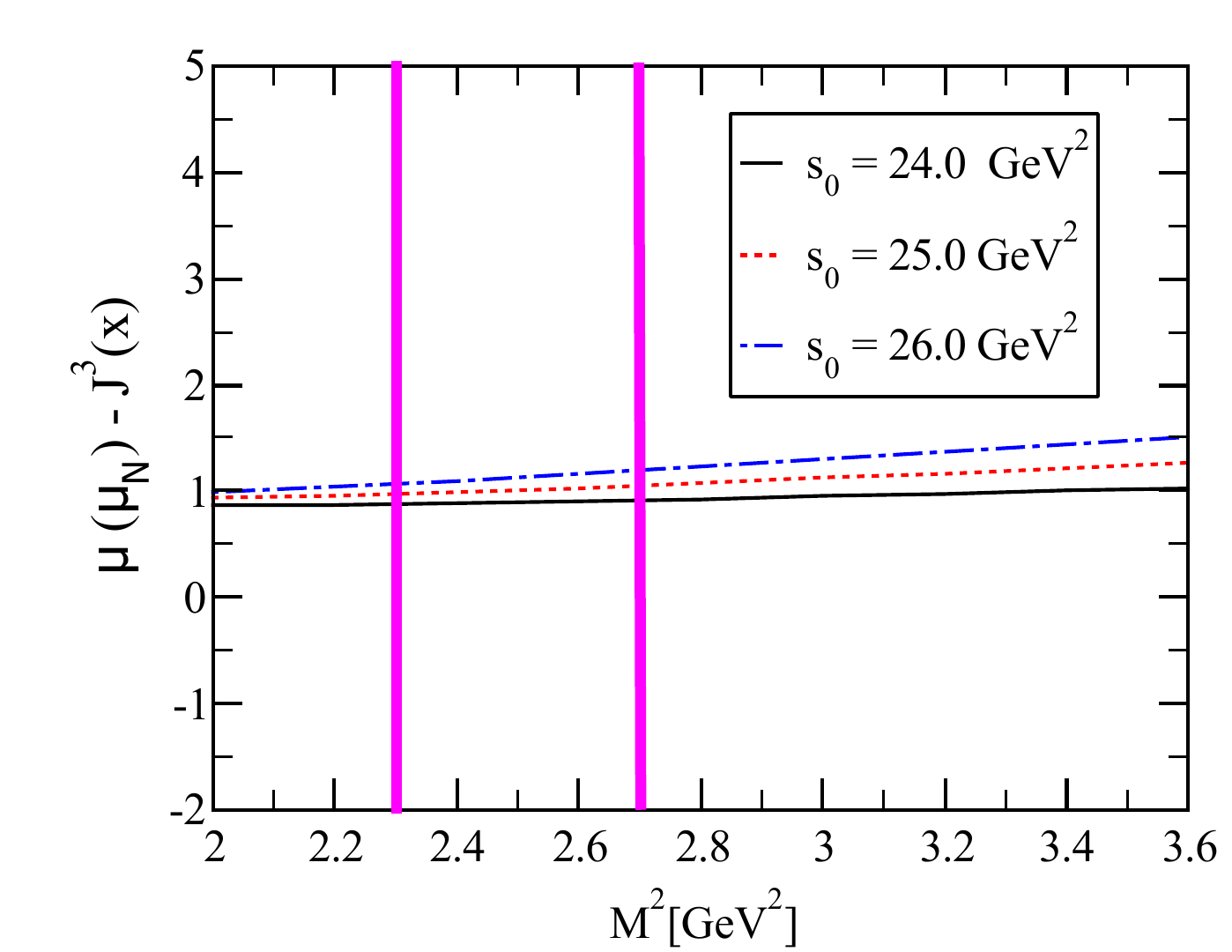}\\
\vspace{0.5 cm}
\includegraphics[width=0.32\textwidth]{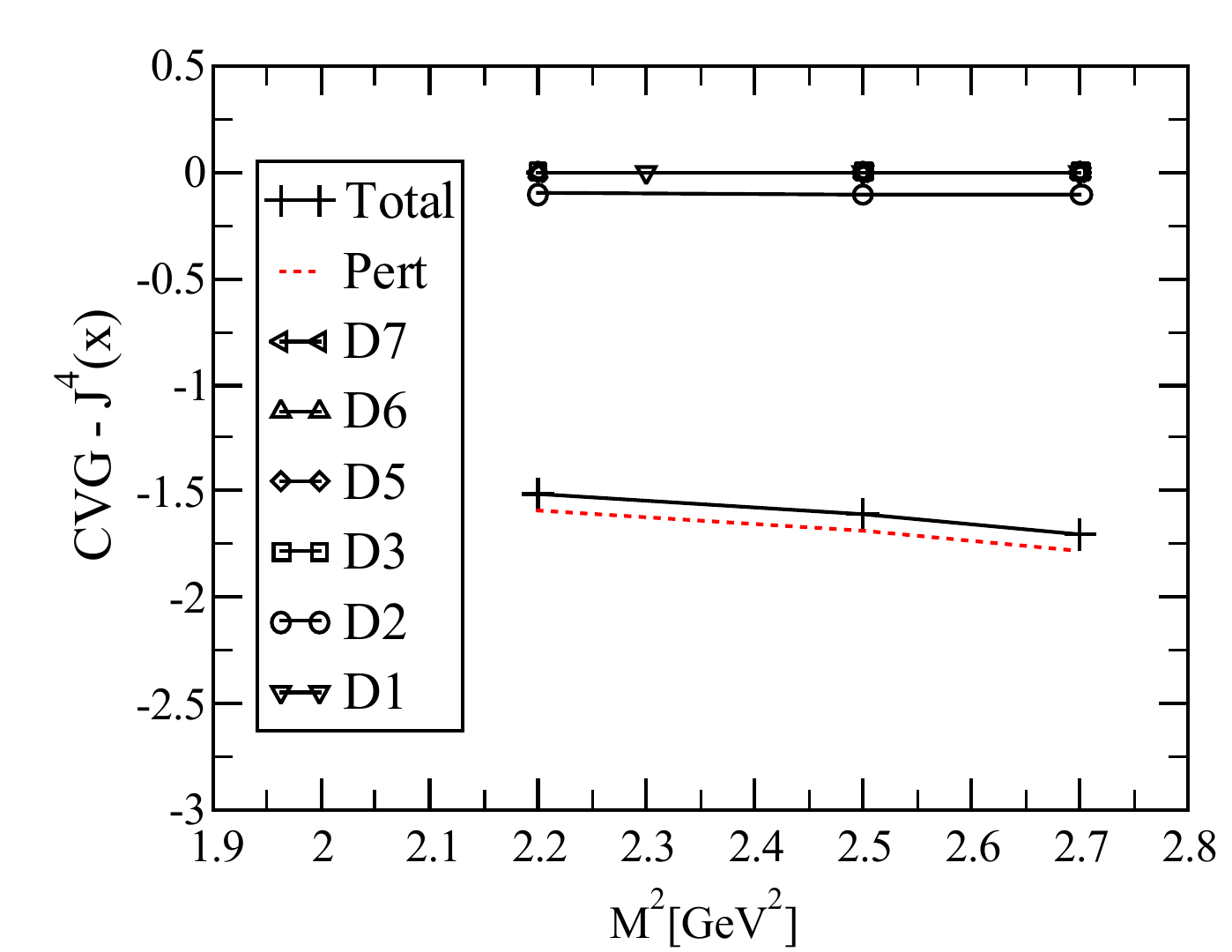}~~
\includegraphics[width=0.32\textwidth]{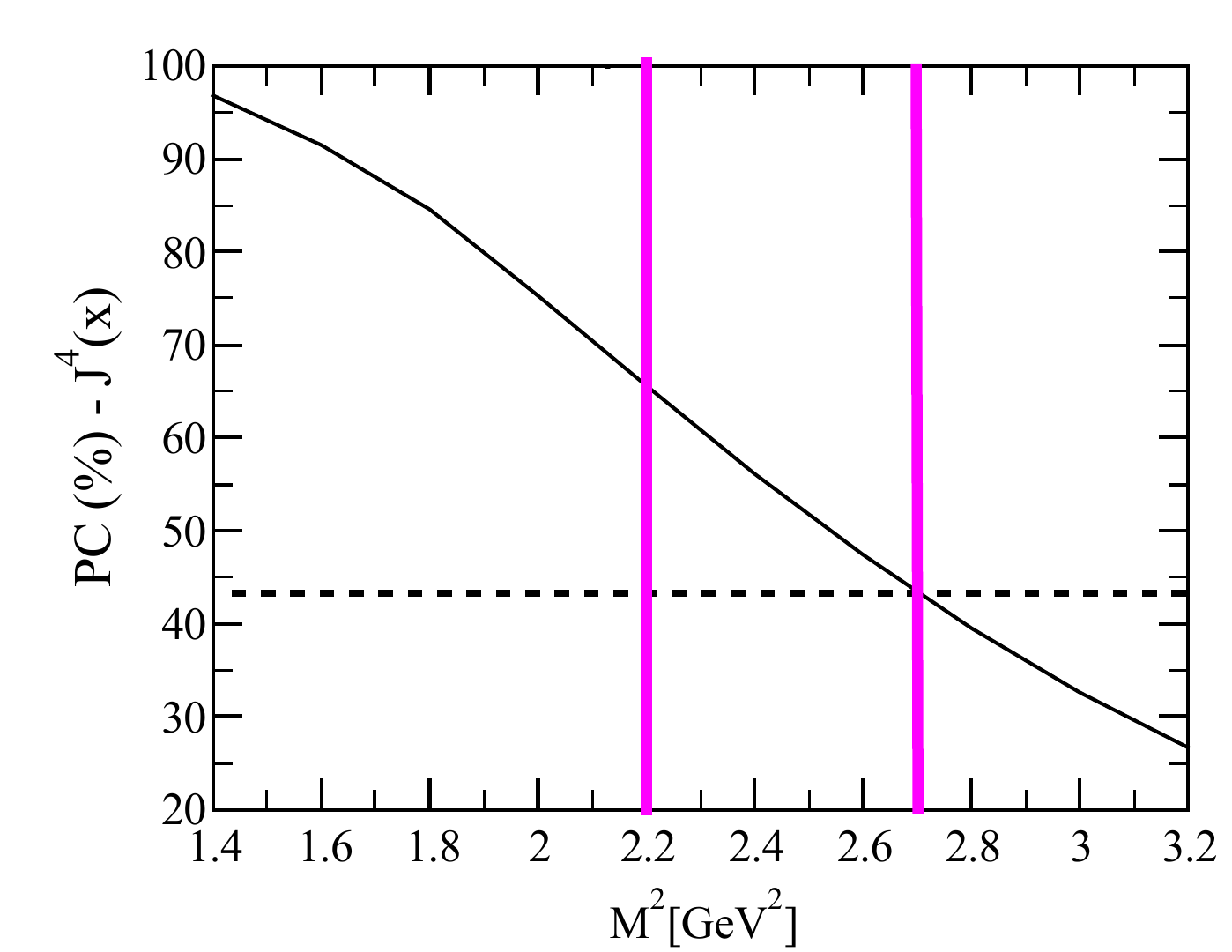}~~
\includegraphics[width=0.32\textwidth]{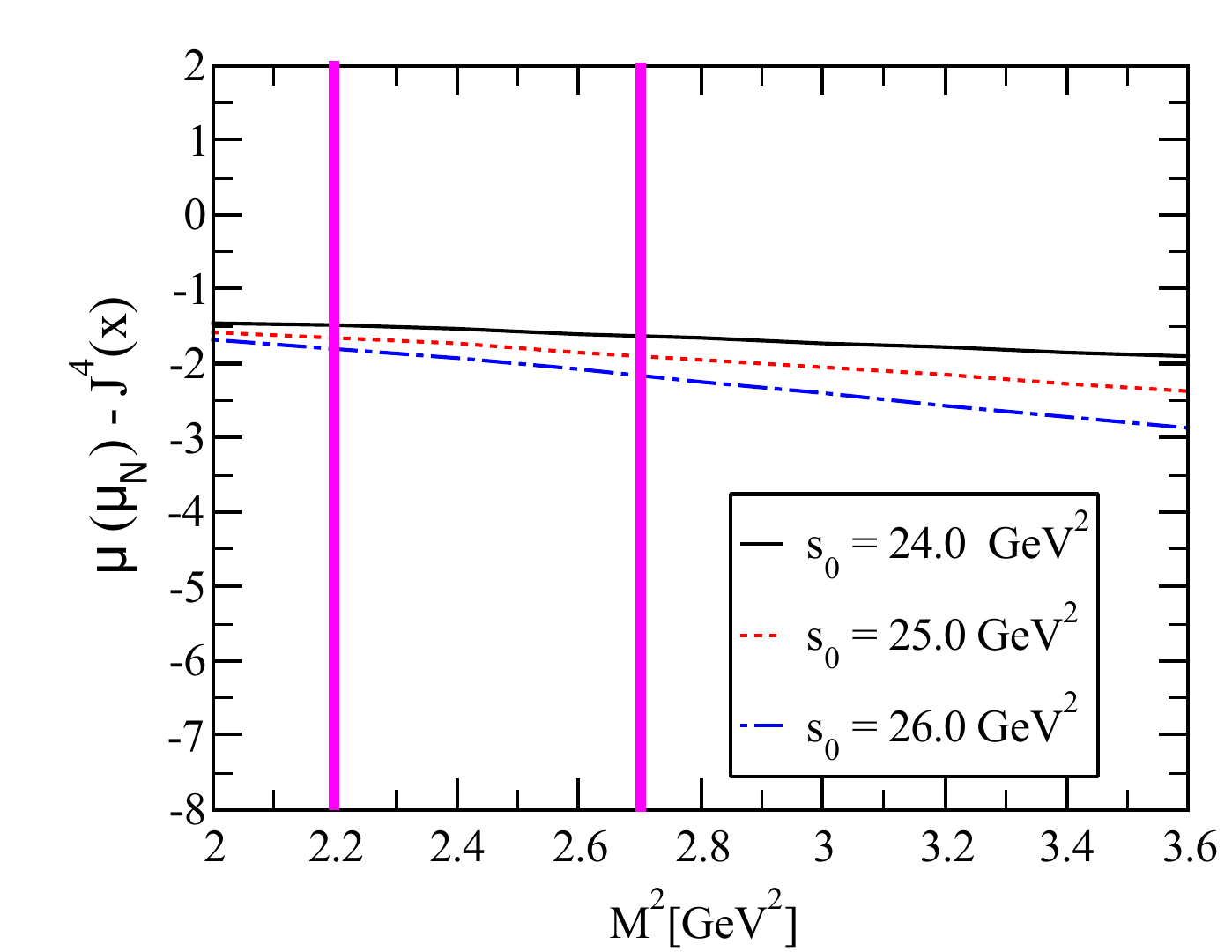}\\
\caption{CVG analysis (left panels),  PC analysis  (middle panels), and total value (right panels) for the magnetic dipole moments of the $P_{\psi s}^{\Lambda}$ pentaquarks versus $\rm{M^2}$ at fixed $\rm{s_0}$ values. In the middle and right panels, the region enclosed by the vertical lines represents the adopted Borel window. The horizontal line in the middle panel corresponds to the smallest PC value obtained within this Borel region in the present study.}
 \label{Msqfig1}
  \end{figure}

The analysis requires two auxiliary quantities: the Borel mass parameter $\rm{M^2}$ and the continuum threshold $\rm{s_0}$. These are fixed by applying the standard procedures and stability criteria intrinsic to the QCD sum rule approach. The acceptable range of $\rm{M^2}$ is determined by requiring proper convergence of the operator product expansion (CVG) and ensuring that the ground-state pole contribution (PC) remains dominant over the continuum. The continuum threshold $\rm{s_0}$ is then chosen such that the extracted hadron mass exhibits minimal sensitivity to variations in $\rm{M^2}$. These conditions can be expressed as
\begin{align}
 \text{PC} &= \frac{\rho_i (\rm{M^2},\rm{s_0})}{\rho_i (\rm{M^2},\infty)} >40\%,
 \qquad 
 \text{CVG}  = \frac{\rho_i^{\text{DimN}} (\rm{M^2},\rm{s_0})}{\rho_i (\rm{M^2},\rm{s_0})} < 1\%,
\end{align}
where $\rho_i^{\text{DimN}} (\rm{M^2},s_0)$ denotes the highest-dimensional terms in the operator product expansion of $\rho_i (\rm{M^2},s_0)$. On the QCD side, the highest-order contributions originate from the dimension-7 operators, $\langle g_s^2 G^2 \rangle \langle \bar q q \rangle$ and 
$\langle g_s^2 G^2 \rangle \langle \bar s s \rangle$. The CVG analysis has therefore been carried out including these D7 terms, and the results presented in Table~\ref{parameter} reflect this choice. In addition, the QCD representation incorporates operator structures such as 
$\langle g_s^2 G^2 \rangle f_{3\gamma}$ (D6), 
$\langle g_s^2 G^2 \rangle \langle \bar q q \rangle \chi$ and 
$\langle g_s^2 G^2 \rangle \langle \bar s s \rangle \chi$ (D5), 
$\langle \bar q q \rangle$ and $\langle \bar s s \rangle$ (D3), 
$f_{3\gamma}$ (D2), and $\langle \bar q q \rangle \chi$ and $\langle \bar s s \rangle \chi$ (D1). Figure~\ref{Msqfig1} shows that, within the chosen Borel window, the PC remains significantly larger than the continuum contribution, thereby ensuring ground-state dominance. The relative magnitudes of the condensate terms confirm the satisfactory convergence of the OPE expansion. Fig.~\ref{Msqfig1} also illustrates the dependence of the magnetic dipole moment of the $P_{\psi s}^{\Lambda}$ pentaquark on the Borel mass parameter $\rm{M^2}$ and the continuum threshold $\rm{s_0}$: only mild variations are observed within the selected parameter ranges.

\begin{table}[htb!]
	\addtolength{\tabcolsep}{10pt}
\caption{Working intervals of $\mathrm{s_0}$ and $\mathrm{M^2}$, fixed by the CVG and PC criteria, together with the magnetic moments obtained for each of the four interpolating currents $J_{1}(x)$--$J_{4}(x)$. The numerical results are presented as predictions associated with the individual currents; their assignment to the observed $P_{\psi s}^{\Lambda}(4338)$ and $P_{\psi s}^{\Lambda}(4459)$ resonances, discussed in Sec.~\ref{formalism}, is adopted as a working hypothesis based on mass predictions and does not enter the construction of this table.}
	\label{parameter}
	\begin{ruledtabular}
\begin{tabular}{l|cccccc}
\\
Current &$\rm{s_0}$ (GeV$^2$) & $\rm{M^2}$ (GeV$^2$)&  ~~  PC ($\%$) ~~ & ~~  CVG ($\%$) & $\mu (\mu_N)$\\
\\
 \hline\hline
                                        \\
$J_1(x)$ & $26.0-28.0$ & $2.4-3.0$ & $67.16-42.56$ &  $ 0.10$  & $-1.35^{+0.35}_{-0.28}$
                    \\
                    \\
$J_2(x)$ & $26.0-28.0$ & $2.4-3.0$ & $67.72-43.26$ &  $ 0.12$  & $~~3.14^{+0.65}_{-0.50}$
                    \\
                    \\
$J_3(x)$ & $24.0-26.0$ & $2.3-2.7$ & $64.48-46.24$ &  $ 0.17$  & $~~1.01^{+0.25}_{-0.20}$
                       \\
                       \\
$J_4(x)$ & $24.0-26.0$ & $2.2-2.7$ & $66.91-44.10$ &  $0.10$  & $-1.79^{+0.41}_{-0.34}$ 
                       \\
                       \\
 \end{tabular}
\end{ruledtabular}
\end{table}
  
The final values of the magnetic moments associated with the four interpolating currents are collected in Table~\ref{parameter}. The reported uncertainties account for variations in the input parameters, the auxiliary quantities $\mathrm{s_0}$ and $\mathrm{M^2}$, and the parameters related to the photon DAs. We emphasize that these four values are best interpreted as predictions for four distinct compact-diquark configurations characterized by the interpolating currents $J_{1}(x)$--$J_{4}(x)$, each probing a different spin--color organization of the constituent diquarks. The mass-based assignment $\{J_{3}(x),J_{4}(x)\}\leftrightarrow P_{\psi s}^{\Lambda}(4338)$ and $\{J_{1}(x),J_{2}(x)\}\leftrightarrow P_{\psi s}^{\Lambda}(4459)$ adopted in Sec.~\ref{formalism} is a working hypothesis: the $\pm 0.11\,\mathrm{GeV}$ mass uncertainties accommodate either physical state within $1\sigma$ of all four currents, and a future measurement of the magnetic moment of either resonance will simultaneously test the compact-diquark hypothesis and the specific current-to-state pairing assumed here. All four values populate the regime $|\mu|\sim 1$--$3\,\mu_N$, which is of the order of a few nuclear magnetons and, as we shall see, lies systematically above the magnitudes predicted by quark-model and heavy pentaquark chiral perturbation theory analyses. Before entering the physical interpretation of the four numerical values, however, we turn to what we regard as the principal result of the present analysis.

\begin{table}[htb!]
\addtolength{\tabcolsep}{6pt}
\caption{Flavor-decomposed magnetic moment contributions for the $P_{\psi s}^{\Lambda}$ states (in units of nuclear magneton $\mu_N$). Here $\mu_q = \mu_u + \mu_d + \mu_s$. The last two columns show the fractional contributions of the light and charm sectors to the total magnetic moment, i.e., $\mu_q/\mu_{\rm tot}$ and $\mu_c/\mu_{\rm tot}$.}
\label{table3}
\begin{ruledtabular}
\begin{tabular}{l|ccccccc}
\\
Current & ~~$\mu_{u}$ & ~~$\mu_{d}$ & ~~$\mu_{s}$ & ~~$\mu_{c}$ & ~~$\mu_{\text{tot}}$ & $\mu_{q}/\mu_{\text{tot}}$ & $\mu_{c}/\mu_{\text{tot}}$ \\
\\
\hline\hline
\\
$J_1(x)$ & $-0.040^{+0.004}_{-0.006}$ & $~~0.020^{+0.003}_{-0.002}$ & $-0.020^{+0.002}_{-0.003}$ & $-1.31^{+0.345}_{-0.275}$ & $-1.35^{+0.35}_{-0.28}$ & $0.03$ & $0.97$ \\
\\
$J_2(x)$ & $~~0.026^{+0.040}_{-0.040}$ & $-0.013^{+0.020}_{-0.020}$ & $~~2.13^{+0.43}_{-0.31}$ & $~~1.00^{+0.20}_{-0.17}$ & $~~3.14^{+0.65}_{-0.50}$ & $0.68$ & $0.32$ \\
\\
$J_3(x)$ & $-2.02^{+0.18}_{-0.14}$ & $~~1.01^{+0.07}_{-0.09}$ & $~~2.02^{+0.16}_{-0.13}$ & $~~0.00^{+0.00}_{-0.00}$ & $~~1.01^{+0.25}_{-0.20}$ & $1.00$ & $0.00$ \\
\\
$J_4(x)$ & $-0.024^{+0.040}_{-0.020}$ & $~~0.012^{+0.010}_{-0.020}$ & $-0.09^{+0.02}_{-0.01}$ & $-1.69^{+0.37}_{-0.32}$ & $-1.79^{+0.41}_{-0.34}$ & $0.06$ & $0.94$ \\
\\
\end{tabular}
\end{ruledtabular}
\end{table}


\subsection{Analytic signatures from the flavor decomposition}\label{subsec:analytic}

\begin{figure}[htb!]
\centering
\includegraphics[width=1.0\textwidth]{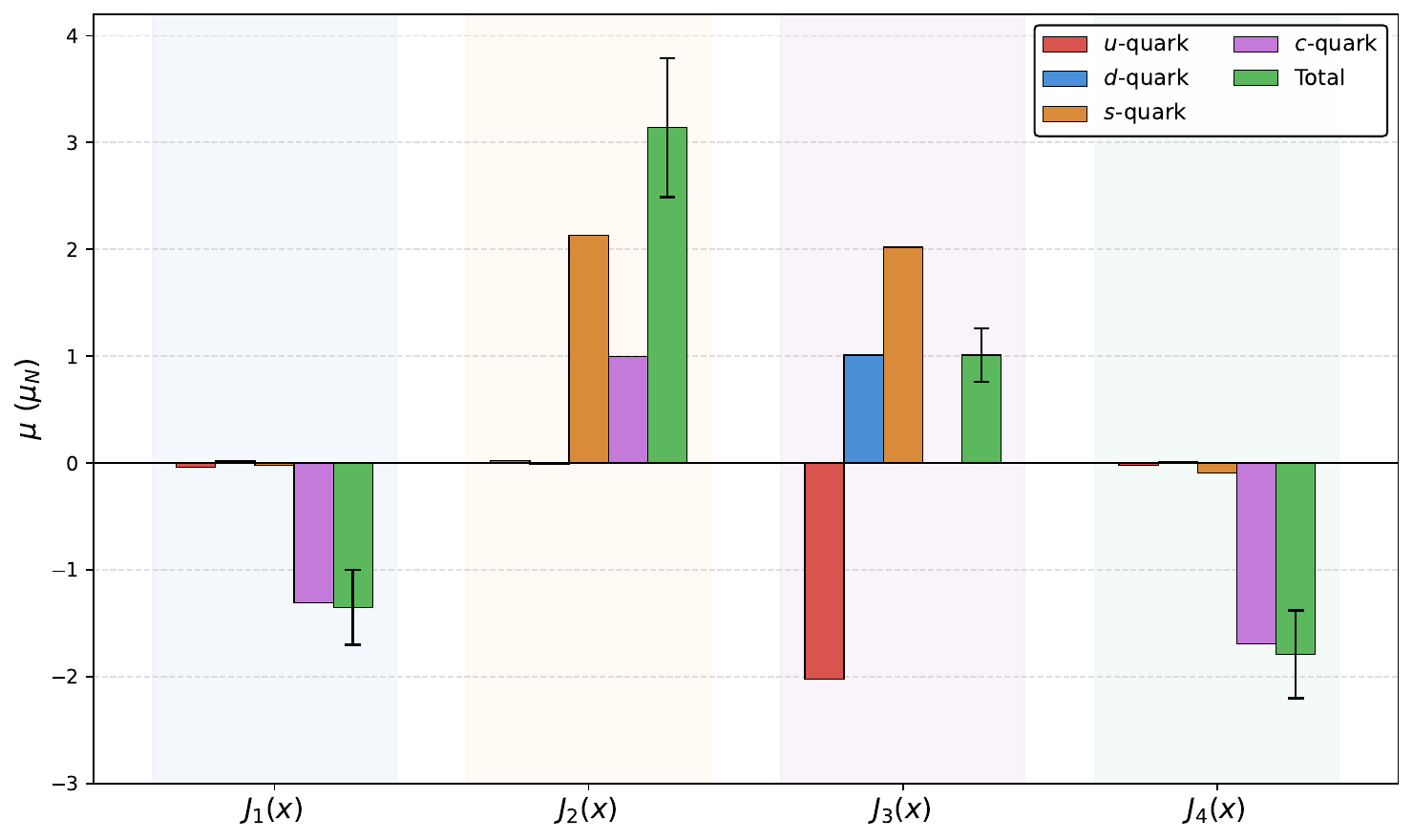}
\caption{Flavor-decomposed contributions to the magnetic moment of the $P_{\psi s}^{\Lambda}$ pentaquark for the four interpolating currents $J_{1}(x)$--$J_{4}(x)$. The individual bars show the $u$-, $d$-, $s$-, and $c$-quark contributions $\mu_{u}$, $\mu_{d}$, $\mu_{s}$, and $\mu_{c}$, while the last bar in each group shows the total magnetic moment $\mu_{\rm tot}$, with the error bar reflecting the combined uncertainty from the input parameters, auxiliary quantities, and photon DAs (see Table~\ref{parameter}). Two analytic signatures discussed in the text are manifest at a glance: (i) in every current the $u$-quark bar is exactly twice the $d$-quark bar with opposite sign, $\mu_{u}/\mu_{d}=-2$; and (ii) for $J_{3}(x)$ the charm-quark bar is absent, $\mu_{c}=0$, whereas for the other three currents the charm sector contributes appreciably (dominantly in $J_{1}(x)$ and $J_{4}(x)$, and coherently with the strange sector in $J_{2}(x)$).}
\label{fig:flavordecomp}
\end{figure}

To expose the microscopic origin of the magnetic moments, we decompose the total results into contributions from individual quark sectors, as summarized in Table~\ref{table3} and visualized in Fig.~\ref{fig:flavordecomp}. This decomposition is obtained by selectively adjusting the charge factors ($e_q$ and $e_c$) within the LCSR, which are explicitly retained for this purpose. Setting $e_c = 0$ eliminates all terms proportional to the charm-quark charge and thereby isolates the light-quark ($u$, $d$, and $s$) contributions in Eqs.~(\ref{F1sonuc})--(\ref{F4sonuc}); repeating the exercise for each light flavor separately yields the individual sector contributions $\mu_u$, $\mu_d$, $\mu_s$, and $\mu_c$ listed in the table and shown as colored bars in the figure. Note that the fractional contributions $\mu_q/\mu_{\rm tot}$ and $\mu_c/\mu_{\rm tot}$ represent the relative weight of each sector in the total moment; when light and charm contributions have opposite signs (e.g., $J_1(x)$ and $J_4(x)$) the fractions can exceed unity in absolute value, reflecting the competing nature of the contributions.

A glance at Fig.~\ref{fig:flavordecomp} reveals two structural regularities that persist across all four currents and that, as we argue below, cannot be attributed to numerical accidents of the OPE. The first is the rigid proportionality between the $u$- and $d$-quark bars in every current, always with opposite sign and with the $u$-quark bar twice the height of the $d$-quark bar. The second is the complete absence of a charm-quark bar in $J_3(x)$, in stark contrast to the sizable charm contributions that dominate $J_1(x)$ and $J_4(x)$ and that add coherently to the strange sector in $J_2(x)$. Both features illustrate how the interpolating-current algebra leaves exact imprints on the magnetic moment beyond any numerical approximation, and both are independent of the specific mapping between the four currents and the two physical resonances. We discuss each in turn.

\paragraph*{Signature I: the light-quark ratio $\mu_u/\mu_d=-2$.} For every one of the four currents we find $\mu_{u}/\mu_{d}=-2$, which is simply the ratio $e_{u}/e_{d}$ of the up- and down-quark charges. This reflects the fact that the $u$- and $d$-quark propagators enter the OPE through a common Lorentz--color kernel, so that the individual light-flavor contributions factorize into an identical dynamical coefficient times the corresponding quark charge: $\mu_{u} = e_{u}\,\mathcal{C}_{ud}$ and $\mu_{d} = e_{d}\,\mathcal{C}_{ud}$ with the same $\mathcal{C}_{ud}$. Consequently, the combined $u+d$ light-quark contribution is controlled by the combination $e_{u}+e_{d}=+1/3$, and never by the isoscalar-like combination $e_{u}+2 e_{d}=0$ that would produce an exact cancellation. The $u$--$d$ partial cancellation visible in Fig.~\ref{fig:flavordecomp} and Table~\ref{table3} is therefore governed purely by charge factors, not by any accidental numerical coincidence, and provides a sharp test of the flavor organization assumed in Eqs.~(\ref{curpcs2})--(\ref{sondenk}), under the standard one-gluon-exchange diquark hypothesis.

\paragraph*{Signature II: analytic vanishing $\mu_c=0$ for $J_3(x)$.} For the $J_3(x)$ current, the charm-quark contribution is $\mu_{c}=0$ identically, not merely small---a feature immediately evident in Fig.~\ref{fig:flavordecomp} as the complete absence of a charm bar in the $J_3(x)$ group. Direct inspection of the explicit expression for $\rho_{3}(\mathrm{M^2},\mathrm{s_0})$ in Eq.~(\ref{F3sonuc}) reveals the complete absence of $e_{c}$-proportional terms in the OPE, which is the technical origin of this analytic cancellation. It is instructive to compare this behavior with that of $J_1(x)$, which also embeds the charm quark in a pseudoscalar diquark of the form $[\bar{q}\,C\gamma_{5}\,c]$ and yet yields a sizable charm-sector contribution $\mu_c^{J_1}=-1.31\,\mu_N$. The two currents differ in two respects: the flavor content and Lorentz structure of the light diquark (scalar $u$-$d$ with $\gamma_5$ in $J_1$ versus vector $u$-$s$ with $\gamma_\mu$ in $J_3$), and, more prominently, the Dirac structure associated with the anti-charm field in the global current---$J_1(x)$ couples the anti-charm through the simple structure $C\bar{c}^T$, whereas $J_3(x)$ introduces the decorated coupling $\gamma_5\gamma^\mu C\bar{c}^T$. The latter structure, when combined with the Dirac organization of the remaining traces in the OPE, produces the exact cancellation of the charm-sector contribution observed in $\rho_3$. The vanishing of $\mu_c^{J_3}$ is therefore a joint consequence of the diquark embedding and the anti-charm coupling in the global current, not a property of the pseudoscalar diquark alone. A fully analytic derivation tracing this cancellation through the individual OPE contractions would be of independent interest but lies beyond the scope of the present analysis. The total magnetic moment of $J_3(x)$ is thus carried entirely by the light sector, $\mu_{q}/\mu_{\rm tot}=1$, a feature dictated by current algebra rather than by numerical details of the OPE.

These two statements---one pattern-type ($\mu_{u}=-2\mu_{d}$) and one analytic zero ($\mu_{c}=0$ for $J_{3}(x)$)---constitute clean, falsifiable structural predictions of the present compact diquark--diquark--antiquark analysis. They survive any reshuffling of the mass-based mapping between the four currents and the two physical pentaquarks, and they can be compared directly to future experimental extractions of flavor-decomposed magnetic moments or to independent theoretical calculations that retain explicit light-flavor charge dependence. As we shall see momentarily, such a comparison has in fact already been carried out for the molecular LCSR analyses of Refs.~\cite{Ozdem:2022kei,Ozdem:2021ugy}. 

A natural question concerns the extent to which these signatures discriminate among different internal configurations: are they exclusive to the compact diquark--diquark--antiquark picture, or could they also arise in alternative structural scenarios? A careful analysis suggests that each signature, taken in isolation, is in fact a structural consequence of the algebra of the interpolating currents rather than of the compact picture per se. Signature~I, $\mu_{u}/\mu_{d}=-2$, follows from the appearance of $u$- and $d$-quark propagators in a common Lorentz--color kernel within the OPE; whether the same ratio is reproduced in alternative interpolating-current organizations---compact or molecular---depends on the specific placement of the $u$ and $d$ propagators within the OPE and on the Lorentz structure of the light-quark coupling, and is not guaranteed a priori. Similarly, Signature~II ($\mu_{c}=0$ for $J_{3}(x)$) reflects a Dirac-structure cancellation that, in principle, could be reproduced by other currents---compact or molecular---whose anti-charm coupling shares the relevant $\gamma_{5}\gamma^{\mu}C\bar{c}^{T}$ organization combined with the appropriate OPE trace structure. We therefore do not claim mathematical uniqueness for either signature in isolation.

To make this discussion concrete, we have applied the same flavor-decomposition procedure to the author's previous molecular LCSR analyses of the two pentaquark states, Refs.~\cite{Ozdem:2022kei,Ozdem:2021ugy}, in which only the total magnetic moments were originally reported. Within those frameworks, the light-quark contributions yield the ratio $\mu_{u}/\mu_{d}=-1/2=e_{d}/e_{u}$ for both $P_{\psi s}^{\Lambda}(4338)$ and $P_{\psi s}^{\Lambda}(4459)$, in contrast to the value $\mu_{u}/\mu_{d}=-2=e_{u}/e_{d}$ obtained in the present compact analysis for all four interpolating currents $J_{1}(x)$--$J_{4}(x)$. The fact that two structurally distinct molecular currents---each corresponding to a different hadronic-molecule decomposition of one of the two states---yield the same inverse ratio strengthens the conclusion that this behavior is a generic feature of the molecular LCSR realizations considered, rather than an accidental property of a particular current. Signature~I is therefore not reproduced in the specific molecular LCSR currents tested so far, and in this LCSR-internal sense it does serve as a discriminator between the compact and molecular organizations studied here. We emphasize, however, that this comparison is restricted to the LCSR framework with the specific interpolating currents employed in Refs.~\cite{Ozdem:2022kei,Ozdem:2021ugy}; the question of whether other molecular current organizations or other theoretical frameworks  reproduce Signature~I lies beyond the scope of the present study.

What is specific to the present analysis, therefore, is the simultaneous occurrence of these two analytic relations across a coherent set of four interpolating currents that collectively span the relevant compact diquark--diquark--antiquark configurations, the resulting magnitude regime $|\mu|\sim 1$--$3\,\mu_{N}$ obtained across all four currents, and the LCSR-internal contrast with the molecular calculations of Refs.~\cite{Ozdem:2022kei,Ozdem:2021ugy} documented above. This combination---the rigid $\mu_{u}/\mu_{d}=-2$ relation in every compact current (versus $-1/2$ in the molecular analyses), the analytic vanishing $\mu_{c}=0$ in $J_{3}(x)$ specifically, the systematic large-magnitude regime, and the structural taxonomy among the four currents discussed in Sec.~\ref{subsec:current-by-current}---constitutes a coherent fingerprint that characterizes our specific compact-diquark realization within the LCSR framework. An independent calculation reproducing only one of these features would not, on this basis, mimic the compact realization studied here. Conversely, a calculation that reproduces the entire package---the two analytic signatures together with the large-magnitude regime and the structural taxonomy---would be indistinguishable from our compact analysis at the level of magnetic moments and would represent an alternative structural realization with the same electromagnetic phenomenology. We regard the experimental test of this package, rather than of any single signature, as the appropriate falsification criterion of the present analysis.

\subsection{Current-by-current interpretation}\label{subsec:current-by-current}

With the analytic signatures in hand, the remaining pattern in Table~\ref{table3} can be read directly from the diquark spin assignment of each interpolating current. We emphasize that the discussion below interprets the four magnetic moments as predictions associated with the four currents themselves: the eventual identification of which physical resonance each prediction describes is a separate phenomenological question that depends on the current-to-state mapping discussed in Sec.~\ref{formalism}. The internal-structure pattern that emerges, however, is independent of that mapping and is dictated solely by the spin--color organization of the interpolating currents.

The results clearly indicate that the charm-quark contribution dominates in the $J_1(x)$ and $J_4(x)$ configurations, while the light sector governs the $J_2(x)$ and $J_3(x)$ configurations. For the $J_1(x)$ current, both diquarks are of scalar type ($[u^T C \gamma_5 d]$ and $[s^T C \gamma_5 c]$). Here, the charm quark is embedded in a spin-0 diquark, leading to partial spin screening but still yielding a dominant charm contribution, $\mu_c/\mu_{\rm tot} \simeq 0.97$. The light-quark contributions are minor and largely cancel each other (subject to the $\mu_{u}/\mu_{d}=-2$ rule), giving a small $\mu_q/\mu_{\rm tot} \simeq 0.03$. The total magnetic moment associated with this current is $\mu_{J_1} = -1.35^{+0.35}_{-0.28}\,\mu_N$. In the $J_2(x)$ current, a scalar light diquark couples to a vector charm-bearing diquark ($[s^T C \gamma_\mu c]$). This structure allows the charm spin to couple directly to the electromagnetic current, producing constructive interference between the $s$ and $c$ sectors. The light contribution ($\mu_q/\mu_{\rm tot} \simeq 0.68$) is dominated by the strange quark ($\mu_{s}=2.13\,\mu_{N}$), while the charm part ($\mu_c/\mu_{\rm tot} \simeq 0.32$) adds coherently, resulting in the largest total moment among all four configurations, $\mu_{J_2} = +3.14^{+0.65}_{-0.50}\,\mu_N$.

The $J_3(x)$ configuration displays the complementary behavior already discussed in Sec.~\ref{subsec:analytic}: the charm-sector contribution vanishes analytically ($\mu_c=0$) as a consequence of the combined Dirac structure of the charm diquark and the anti-charm coupling in the global current, so that the total magnetic moment $\mu_{J_3}=1.01^{+0.25}_{-0.20}\,\mu_N$ arises entirely from the light sector. Here the $u$- and $d$-quark contributions partially cancel via the $\mu_{u}=-2\mu_{d}$ relation, and the strange-quark contribution $\mu_{s}=+2.02\,\mu_{N}$ combined with the residual $u+d$ piece yields the observed total. Finally, the $J_4(x)$ current involves two vector diquarks, $[u^T C \gamma_\mu s]$ and $[d^T C \gamma^\mu c]$, coupled to the anti-charm quark. In this fully spin-1 configuration, the charm quark contributes overwhelmingly ($|\mu_c/\mu_{\rm tot}|\approx 0.94$), while the light-quark sector provides a small correction ($|\mu_q/\mu_{\rm tot}|\approx 0.06$). The total value, $\mu_{J_4}=-1.79^{+0.41}_{-0.34}\,\mu_N$, is large and negative. The sign reversal relative to $J_2(x)$ originates from the opposite phase alignment of the vector diquarks and the coupling to the anti-charm quark, which flips the direction of the charm contribution relative to the total spin.

Taken together, the four currents define a clear structural taxonomy independent of which physical state each is eventually associated with. Scalar diquarks containing the charm quark ($J_1(x)$, $J_3(x)$) tend to screen or fully suppress the charm contribution, while vector diquarks containing the charm quark ($J_2(x)$, $J_4(x)$) activate it, leading to large absolute values with opposite signs depending on their phase couplings. The electromagnetic properties---particularly the magnetic moment---therefore serve as a powerful diagnostic of the internal spin--flavor dynamics and spatial correlations in multiquark systems.

A general expectation in hadron physics is that physical observables should remain invariant under a change of basis for the hadronic states. For magnetic moments this principle applies only with important qualifications: the observable is intimately connected to the internal spatial and spin organization of the constituent quarks, and adopting a different basis effectively corresponds to examining a distinct configuration of the internal structure, which can lead to substantial variations in the predicted magnetic moments. This sensitivity has been repeatedly observed in studies of tetra- and pentaquark states across a variety of theoretical frameworks---ranging from quark-model and effective-field-theory analyses to LCSR  calculations employing different interpolating current organizations---where the predicted magnetic moments are found to differ significantly in both magnitude and sign depending on the assumed internal configuration~\cite{Wang:2016dzu, Ozdem:2024rch, Gao:2021hmv, Li:2024wxr, Li:2024jlq, Ozdem:2024rqx, Ozdem:2024txt, Mutuk:2024ach, Ozdem:2024dbq, Ozdem:2024lpk, Azizi:2023gzv, Ozdem:2026gwt}.

Far from constituting a weakness of the multi-current strategy, this sensitivity is its principal virtue. Different structural assumptions yield predictably distinct results, and the exact relations among them---the two analytic signatures identified in Sec.~\ref{subsec:analytic}---are enforced by the algebra of the interpolating currents. A future magnetic-moment measurement of either $P_{\psi s}^{\Lambda}$ resonance would therefore not merely test ``the'' compact-diquark hypothesis as a single object: it would specifically identify which of the four spin--flavor organizations encoded in $J_{1}(x)$--$J_{4}(x)$ is realized in the physical state, and would simultaneously test the mass-based phenomenological pairing adopted here. The apparent ambiguity of having four predictions where two physical resonances are observed is thereby transformed into experimental leverage: each compact-diquark configuration carries a distinctive magnetic signature, and a measurement determines which is realized in nature.

\begin{figure}[htb!]
\centering
\includegraphics[width=0.97\textwidth]{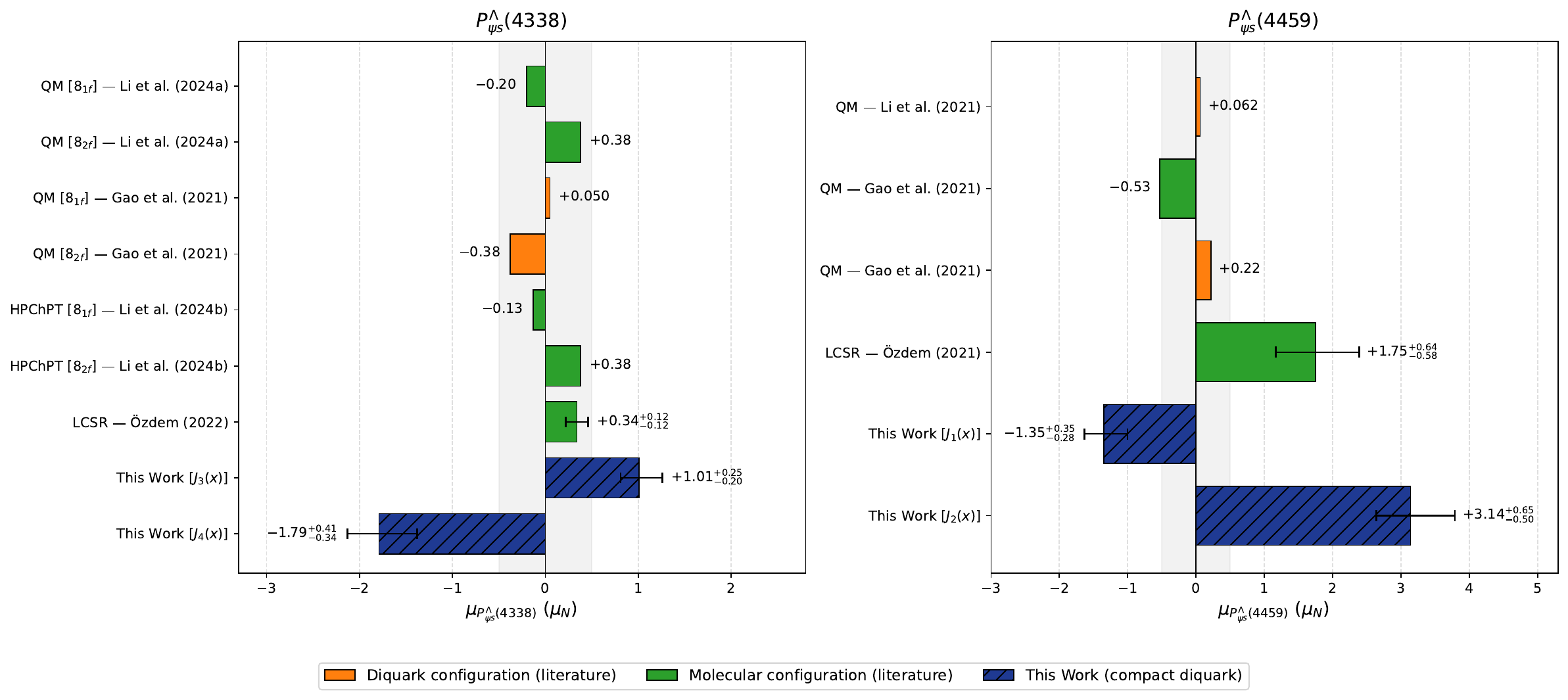}
\caption{Comparison of theoretical predictions for the magnetic 
moment of the $P_{\psi s}^{\Lambda}(4338)$ (left) and 
$P_{\psi s}^{\Lambda}(4459)$ (right) pentaquarks. Orange bars 
denote literature predictions obtained with compact 
diquark--diquark--antiquark configurations, green bars those 
obtained in molecular or hadronic-molecule pictures, and blue 
hatched bars the results of the present work. The shaded gray 
band indicates the small-moment regime $|\mu|\lesssim 0.5\,\mu_N$ 
typical of QM and HPChPT predictions. Literature values 
are taken from Li \textit{et al.}~\cite{Li:2024wxr} [2024a], 
Gao \textit{et al.}~\cite{Gao:2021hmv}, Li \textit{et al.}~\cite{Li:2024jlq} [2024b], 
Özdem~\cite{Ozdem:2022kei}, Li \textit{et al.}~\cite{Li:2021ryu}, 
and Özdem~\cite{Ozdem:2021ugy}. For $\mu_{P_{\psi s}^{\Lambda}(4338)}$, the four QM entries correspond to two distinct flavor 
octet assignments ($8_{1f}$ and $8_{2f}$). The grouping of our predictions by physical resonance follows the mass-based working hypothesis adopted in Sec.~\ref{formalism}; the four current-based predictions $\mu_{J_{1}}$--$\mu_{J_{4}}$ are themselves independent of this grouping.}
\label{fig:comparison}
\end{figure}

\subsection{Comparison with other theoretical approaches}\label{subsec:comp}

Comparison with existing theoretical approaches, illustrated in Fig.~\ref{fig:comparison}, provides valuable insight into the implications of our results. To enable a direct comparison with literature predictions, which are organized by physical resonance, we group our four current-based predictions according to the mass-based pairing adopted in Sec.~\ref{formalism}, while keeping in mind that this pairing is a working hypothesis subject to revision by a future magnetic-moment measurement.

For the $P_{\psi s}^{\Lambda}(4338)$ state, all quark-model (QM)~\cite{Li:2024wxr,Gao:2021hmv} and heavy pentaquark chiral perturbation theory (HPChPT)~\cite{Li:2024jlq} predictions cluster within $|\mu|\lesssim 0.4\,\mu_N$, with substantial sign variations depending on the assumed flavor multiplet representation. It is worth noting in this context that  HPChPT is an effective field theory tailored to loosely bound molecular states near hadronic thresholds, so the small magnitudes it produces are intrinsically consistent with the molecular picture and should not be regarded as an independent validation of any particular structural scenario. The molecular LCSR calculation of~\cite{Ozdem:2022kei} falls in the same small-magnitude regime, consistent with the expectations for a spatially extended, loosely bound hadronic molecule. In contrast, our compact diquark--diquark--antiquark predictions for the currents associated with this state under the working pairing, $\mu_{J_{3}}=1.01^{+0.25}_{-0.20}~\mu_{N}$ and $\mu_{J_{4}}=-1.79^{+0.41}_{-0.34}~\mu_{N}$, lie three to five times above this common range and carry a wider sign variation---features that directly reflect the internal spin organization of the compact scenario rather than any particular flavor multiplet choice.

The situation for the $P_{\psi s}^{\Lambda}(4459)$ is more nuanced and merits explicit comment, since the molecular LCSR analysis of~\cite{Ozdem:2021ugy} yields $\mu=1.75^{+0.64}_{-0.58}~\mu_{N}$---a value of order one nuclear magneton and thus comparable in magnitude to our compact-diquark predictions for the currents associated with this state under the working pairing, $\mu_{J_{1}}=-1.35^{+0.35}_{-0.28}~\mu_{N}$ and $\mu_{J_{2}}=+3.14^{+0.65}_{-0.50}~\mu_{N}$. QM  estimates~\cite{Li:2021ryu,Gao:2021hmv} again predict modest magnitudes, $|\mu|\lesssim 0.5~\mu_N$, but the existence of a molecular LCSR prediction in the same magnitude regime as our compact results constitutes a genuine challenge to any simple magnitude-based discrimination between compact and molecular configurations. We acknowledge this point openly: a total-moment measurement alone, even if it returns a value of order $1\,\mu_N$, cannot cleanly separate ``compact'' from ``molecular'' within the LCSR framework, since both types of interpolating currents couple to specific spin--color structures and yield results sensitive to those choices. What remains robust at the qualitative level is the pattern visible in Fig.~\ref{fig:comparison}: compact diquark--diquark--antiquark configurations systematically populate the region $|\mu|\gtrsim 1~\mu_{N}$ with a wide sign variation, whereas QM and HPChPT approaches cluster at much smaller magnitudes, $|\mu|\lesssim 0.5~\mu_{N}$. This regime separation is, importantly, independent of which specific current--state pairing is adopted: any reshuffling of the mass-based mapping would simply reassign our four predictions among the two resonances but leave the overall regime structure intact.

When a measurement becomes available, these two distinct regimes will play complementary roles. A small measured magnitude would strongly favor the QM/HPChPT picture and disfavor the compact diquark--diquark--antiquark description. A large magnitude would be consistent with both our compact scenario and with specific molecular LCSR implementations such as that of~\cite{Ozdem:2021ugy}; as shown in Sec.~\ref{subsec:analytic}, the two analytic signatures---the $\mu_u/\mu_d=-2$ relation and the vanishing $\mu_c=0$ for $J_3(x)$---already distinguish the compact and molecular pictures within the LCSR framework itself at the flavor-decomposed level, with the molecular currents of Refs.~\cite{Ozdem:2022kei,Ozdem:2021ugy} yielding $\mu_u/\mu_d=-1/2$ rather than $-2$. A measurement would thus simultaneously test three propositions: (i) the compact diquark--diquark--antiquark hypothesis as a whole, (ii) the specific spin--flavor organization encoded in one of the four currents $J_{1}(x)$--$J_{4}(x)$, and (iii) the mass-based current-to-state pairing assumed in this work, with the additional benefit that the flavor-decomposed comparison documented in Sec.~\ref{subsec:analytic} already provides an LCSR-internal discriminator between the compact and molecular realizations independent of the total magnitude. 

In light of the magnitude overlap with Ref.~\cite{Ozdem:2021ugy}, the value added by the present multi-current analysis lies precisely in the structural information that goes beyond a single total-moment prediction. Three features of our analysis, taken together, address the limitation acknowledged above and provide diagnostic content that a single molecular LCSR calculation cannot offer. First, the two analytic signatures of Sec.~\ref{subsec:analytic}---the rigid $\mu_{u}/\mu_{d}=-2$ relation across all four currents and the analytic vanishing $\mu_{c}=0$ for $J_{3}(x)$---are flavor-decomposed predictions whose experimental or theoretical verification provides a structural test independent of the total magnitude. In particular, the LCSR-internal comparison documented in Sec.~\ref{subsec:analytic} shows that Signature~I is not reproduced in the molecular currents of Refs.~\cite{Ozdem:2022kei,Ozdem:2021ugy}, where the same flavor-decomposition procedure yields $\mu_{u}/\mu_{d}=-1/2$ instead, providing a concrete diagnostic example of how flavor-decomposed observables can discriminate between compact and molecular LCSR realizations even when the total magnitudes are comparable. Second, the systematic appearance of $|\mu|\sim 1$--$3\,\mu_{N}$ across all four currents---rather than only the single value $\mu\simeq 1.75\,\mu_N$ obtained in the molecular LCSR analysis of Ref.~\cite{Ozdem:2021ugy}---reveals the regime structure of the compact diquark--diquark--antiquark family as a whole, not the magnetic response of one particular interpolating current. Third, the structural taxonomy among the four currents discussed in Sec.~\ref{subsec:current-by-current} (scalar-charm versus vector-charm diquarks, $\mu_c$-dominated versus light-dominated regimes) provides a current-by-current diagnostic of the internal spin--color organization that is absent from any single-current calculation. We therefore regard the experimental test of this composite structural package---rather than of the total magnetic moment of a single state in isolation---as the appropriate falsification criterion of the present analysis. More broadly, the multi-current strategy with flavor-decomposed observables illustrated here offers a diagnostic template that, in principle, can be applied to other compact multiquark candidates whose internal organization is similarly under debate, though such extensions lie beyond the scope of the present study.

\subsection{Experimental Outlook}\label{sec:outlook}

Although a direct measurement of the magnetic moment of short-lived states such as the $P_{\psi s}^{\Lambda}$ pentaquarks remains challenging, it is not beyond the reach of future experiments. Several complementary strategies can be envisaged to constrain, and eventually measure, this fundamental observable.

The most promising approach exploits the fact that the magnetic moment governs, together with other electromagnetic form factors, the coupling strength of the pentaquark to the electromagnetic field. The photoproduction rate in reactions of the type $ep\to e'P_{\psi s}^{\Lambda}X$ near threshold is sensitive to this coupling, and a precise determination of the corresponding cross section---interpreted within a theoretical framework that incorporates the magnetic moment as an explicit parameter---would yield the first quantitative constraints on this quantity. Complementary information can be extracted from the angular distributions of the decay products $P_{\psi s}^{\Lambda}\to J/\psi\Lambda$ in photoproduction events, which carry imprints of spin-dependent interactions. The use of polarized photon beams would significantly enhance the sensitivity to these effects and provide an independent probe of the magnetic-moment-driven dynamics. Radiative decays such as $P_{\psi s}^{\Lambda}\to J/\psi\Lambda\gamma$ probe related electromagnetic transition couplings, although their exploitation is hampered by the very small expected branching ratios.

From an experimental standpoint, a precise measurement requires high luminosity, excellent vertexing and tracking capabilities, and good photon energy resolution. Facilities such as the upgraded LHCb, Belle II, and the future Electron--Ion Collider are well suited to pursue these measurements, although the short lifetime of the $P_{\psi s}^{\Lambda}$ states, $\tau\lesssim 10^{-23}$~s, and their correspondingly low production rates pose significant challenges that demand very high integrated luminosities. Ultra-peripheral heavy-ion collisions offer an alternative high-photon-flux environment, and electron--ion colliders could, in principle, contribute to mapping the $Q^{2}$ behavior of the electromagnetic form factors, providing independent cross-checks of the underlying structure.

Beyond the total magnetic moment, the two analytic signatures identified in this work offer a particularly attractive target for theory-driven experimental analyses. A flavor-resolved interpretation of $P_{\psi s}^{\Lambda}$ production and decay---obtained, for instance, by combining measurements of production or polarization observables in channels with different quark-flavor sensitivities---could provide indirect constraints on the flavor-decomposed contributions, including the relation $\mu_{u}/\mu_{d}=-2$ and, in appropriate channels, the vanishing of the charm-quark contribution for $J_3(x)$-type currents. Such constraints would constitute experimental checks of the compact diquark--diquark--antiquark picture that rely on relations among observables rather than on the absolute magnitude of the total moment alone, and would therefore be complementary to a single-moment measurement. A coordinated theoretical and experimental program combining photoproduction measurements, polarization observables, radiative-decay studies, and form-factor mappings would transform the magnetic moment from a theoretical parameter into a key observable for deciphering the internal structure of the $P_{\psi s}^{\Lambda}$ pentaquarks. We therefore strongly encourage future high-luminosity experiments to pursue this measurement as a crucial test of exotic hadron structure, while acknowledging the significant experimental challenges involved.

\section{Summary}\label{summary}

The principal message of this work is methodological: the multi-current LCSR framework produces, in addition to standard numerical predictions for electromagnetic observables, exact analytic relations that encode structural information about compact multiquark configurations. We have demonstrated this explicitly for the hidden-charm strange pentaquarks $P_{\psi s}^{\Lambda}(4338)$ and $P_{\psi s}^{\Lambda}(4459)$, under the working assumption $J^{P}=\tfrac{1}{2}^{-}$ for both states, by constructing four distinct diquark--diquark--antiquark interpolating currents and extracting both the total magnetic moments and their flavor-sector decomposition.

Two analytic signatures emerged from the flavor decomposition that, in our view, constitute the central result of the present analysis. First, in all four currents the light-quark contributions satisfy the exact ratio $\mu_{u}/\mu_{d}=e_{u}/e_{d}=-2$, which shows that $u$ and $d$ enter the QCD side through a common dynamical kernel and contribute only through their electric charges. Second, for the $J_{3}(x)$ current the charm-quark contribution vanishes identically, $\mu_{c}=0$. This cancellation cannot be attributed to the pseudoscalar embedding of the charm diquark alone, since the same embedding occurs in $J_{1}(x)$ and yet produces a sizable $\mu_{c}^{J_{1}}=-1.31\,\mu_{N}$; the distinguishing feature is rather the Dirac structure of the anti-charm coupling in the global current, which combines with the remaining OPE traces to enforce the exact vanishing of $e_{c}$-proportional terms in $\rho_{3}$. Both signatures are enforced by the algebra of the compact interpolating currents themselves and are therefore independent of any phenomenological mapping between the four currents and the two physical resonances; the LCSR-internal comparison documented in Sec.~\ref{numerical} further demonstrates that this algebraic structure distinguishes the compact realization from the molecular currents of Refs.~\cite{Ozdem:2022kei,Ozdem:2021ugy} at the flavor-decomposed level. As such, the signatures provide clean, falsifiable structural predictions of the present analysis that survive the unavoidable ambiguities of state-to-current assignments, and that, when combined with the magnitude regime and structural taxonomy discussed in Sec.~\ref{numerical}, form the falsifiable package characterizing our specific compact-diquark realization. 

The numerical values of the magnetic moments---to be read as predictions associated with the four interpolating currents rather than as definitive predictions for the two observed resonances---are $\mu_{J_{1}}=-1.35^{+0.35}_{-0.28}\,\mu_{N}$, $\mu_{J_{2}}=+3.14^{+0.65}_{-0.50}\,\mu_{N}$, $\mu_{J_{3}}=1.01^{+0.25}_{-0.20}\,\mu_{N}$, and $\mu_{J_{4}}=-1.79^{+0.41}_{-0.34}\,\mu_{N}$. The mass-based assignment adopted here---pairing $\{J_{3}(x),J_{4}(x)\}$ with the $P_{\psi s}^{\Lambda}(4338)$ and $\{J_{1}(x),J_{2}(x)\}$ with the $P_{\psi s}^{\Lambda}(4459)$ on the basis of the mass predictions of~\cite{Wang:2025fqh}---is adopted as a working hypothesis rather than as a definitive structural identification. The $\pm 0.11~\text{GeV}$ mass uncertainties overlap both physical states at the $1\sigma$ level, and a future measurement of the magnetic moment will therefore simultaneously test the compact-diquark hypothesis, the specific spin--flavor organization encoded in one of the four currents, and the pairing of currents with observed resonances assumed here.

The sign reversals and factor-of-two magnitude variations among the four currents highlight the deep sensitivity of the magnetic moment to the internal organization of each interpolating current: in the $J_{3}(x)$ configuration the charm-sector contribution cancels analytically due to the combined Dirac structure of the charm diquark and the anti-charm coupling, whereas in $J_{4}(x)$ a vector charm-bearing diquark $[d^T C\gamma^\mu c]$ generates a dominant, negative charm contribution. An analogous structural pattern drives the difference between the $J_{1}(x)$ and $J_{2}(x)$ results. The analytic signatures identified above, by contrast, are immune to these structural variations and constitute the most robust predictions of the present analysis.

When compared with previous theoretical approaches, our compact-diquark predictions stand out collectively. QM and HPChPT predictions typically lie in the range $|\mu|\lesssim 0.5\,\mu_{N}$, whereas our four diquark--diquark--antiquark configurations populate the region $|\mu|\sim 1$--$3\,\mu_{N}$. Comparison with other LCSR-based studies is more subtle: the molecular LCSR analysis of~\cite{Ozdem:2022kei} for the $P_{\psi s}^{\Lambda}(4338)$ falls in the small-moment range, whereas that of~\cite{Ozdem:2021ugy} for the $P_{\psi s}^{\Lambda}(4459)$ produces a magnitude comparable to our compact predictions. This mixed pattern implies that the total magnetic moment is most effective as a discriminator when interpreted alongside the methodological assumptions of each calculation, and it underscores the added value of the two analytic signatures identified here. Specifically, applying the same flavor-decomposition procedure to the molecular LCSR analyses of Refs.~\cite{Ozdem:2022kei,Ozdem:2021ugy} yields $\mu_{u}/\mu_{d}=-1/2$ for both states, in contrast to the value $-2$ obtained here for all four compact currents, providing an LCSR-internal discriminator that operates at the level of flavor-decomposed contributions even when the total magnitudes are comparable.

Overall, this analysis underscores that magnetic moments and their flavor decomposition encode detailed and partly universal information about the spin correlations and color dynamics within multiquark systems. More broadly, it illustrates that the multi-current LCSR strategy is not merely a computational device but a diagnostic framework in which the algebra of interpolating currents enforces exact relations among physical observables---relations that can serve as model-independent structural predictions for exotic multiquark states. This work, together with earlier studies on various exotic states~\cite{Ozdem:2025jda, Ozdem:2024rch, Ozdem:2024rqx, Ozdem:2023htj, Ozdem:2022kei, Wang:2016dzu, Ortiz-Pacheco:2018ccl, Xu:2020flp, Ozdem:2018qeh, Ozdem:2021ugy, Li:2021ryu, Gao:2021hmv, Guo:2023fih, Wang:2022nqs, Wang:2022tib, Ozdem:2024jty, Li:2024wxr, Li:2024jlq, Mutuk:2024ltc, Mutuk:2024jxf, Mutuk:2024ach, Ozdem:2024usw, Ozdem:2025fks, Zhu:2025abk, Ozdem:2026gmn, Ozdem:2025ion, Mutuk:2026zxp}, contributes to the development of such a framework.

\clearpage

\appendix

\begin{widetext}

\section{Explicit form of the spectral densities $\rho_i(\mathrm{M^2},\mathrm{s_0})$}\label{appa}
 
The explicit form of the functions $\rho_1 (\mathrm{M^2},\mathrm{s_0})$, $\rho_2 (\mathrm{M^2},\mathrm{s_0})$, $\rho_3 (\mathrm{M^2},\mathrm{s_0})$, and $\rho_4 (\mathrm{M^2},\mathrm{s_0})$, obtained following the implementation of all the procedures described in Sec.~\ref{formalism}, are presented in Eqs.~(\ref{F1sonuc})--(\ref{F4sonuc}) below. Before quoting the explicit expressions, it is useful to outline their general organization for the reader's convenience.
 
Each spectral density admits a natural decomposition by the origin of its individual terms. The terms proportional to $I[0,n]$ alone (no condensate or DA prefactor) constitute the leading perturbative contribution arising from short-distance photon emission off the quark lines, as implemented through the replacement of Eq.~(\ref{free}). Terms multiplied by the quark condensates $\langle\bar{q}q\rangle$ or $\langle\bar{s}s\rangle$ encode the lowest-dimensional non-perturbative contributions associated with chiral-symmetry breaking in the QCD vacuum. The gluon condensate $\langle g_s^2 G^2\rangle$ appears either alone or combined with quark condensates, generating higher-dimensional contributions that mix perturbative and non-perturbative gluonic dynamics. The parameter $f_{3\gamma}$, normalization constant of the twist-3 photon DAs, multiplies terms involving the functions $\mathcal{V}(\alpha_i)$, $\mathcal{A}(\alpha_i)$, and $\psi^a(u)$. The magnetic susceptibility $\chi$ multiplies terms involving the leading twist-2 photon DA $\varphi_\gamma(u)$ and parametrizes long-distance photon emission from quark condensates. Finally, the auxiliary functions $I_i[\mathcal{F}]$ (defined at the end of this appendix together with the integration measure $\mathcal{D}\alpha_i$) encapsulate the convolutions of three-particle photon DAs $\mathcal{S}(\alpha_i)$, $\mathcal{T}_i(\alpha_i)$, $\mathbb{A}(u)$, and others with the relevant kinematic structures.
 
The electric charges $e_q$ and $e_c$ are retained explicitly throughout, since their selective adjustment underlies the flavor decomposition discussed in Sec.~\ref{subsec:analytic}. In particular, the complete absence of $e_c$-proportional terms in $\rho_3(\mathrm{M^2},\mathrm{s_0})$ is directly visible from the expression below.
 
The explicit expressions are:
 \begin{align}
\label{F1sonuc}
 \rho_1 (\mathrm{M^2},\mathrm{s_0})&=-\frac {e_c } {2^{26} \times 5^3 \times 7^2 \pi^7} \Big[ 10780 m_c m_s I[0, 6] - 1647 I[0, 7]\Big]\nonumber\\
      &+\frac {e_s \, m_c \langle g_s^2G^2\rangle  \langle \bar ss \rangle} {2^{24} \times 3^6 \times 5 \pi^5} I_3[\mathcal S] I[0, 5]\nonumber\\
       &-\frac { f_{3\gamma}\langle g_s^2G^2\rangle } {2^{29} \times 3^6 \times 5 \pi^5}  \Big[ 207 (4 e_d + 5 e_s) I_1[\mathcal V] I[0, 4] + 828 e_u I_2[\mathcal V] I[0, 4] - 
 32 (44 e_d m_c m_s I[0, 3] - 44 e_u m_c m_s I[0, 3]\nonumber\\
 &- 
    9 e_d I[0, 4] - 8 e_s I[0, 4] + 9 e_u I[0, 4]) \psi^a[u_0] \Big]\nonumber
       \nonumber\\
        &+\frac {e_s\,m_c \langle \bar ss \rangle} {2^{22} \times 5^2 \pi^5}\Big[ I_3[\mathcal S] I[0, 5] \Big]\nonumber\\
       &+\frac {  f_{3\gamma}  } {2^{25} \times 3^2 \times 5^2  \pi^5}  \Big[  m_c m_s (e_d I_1[\mathcal V] + e_u I_2[\mathcal V]) I[0, 5] - 
 5  ((e_d + e_s) I_1[\mathcal V] + e_u I_2[\mathcal V]) I[0, 6] \Big],
 \end{align}
           \begin{align}
 \rho_2 (\mathrm{M^2},\mathrm{s_0})&=\frac {1 } {2^{25}  \times 3^4 \times 5^3  \times 7^2 \pi^7} \Big[ 32620 e_c m_c m_s I[0, 6] + 1647 (2 e_c - 9 e_s) I[0, 7] \Big]\nonumber\\ \label{F2sonuc}
      &+\frac {e_s \, m_c \langle g_s^2G^2\rangle  \langle \bar ss \rangle} {2^{24} \times 3^4 \times 5 \pi^5} \Big [ (\mathbb A[u_0] + 13 I_3[\mathcal S]) I[0, 3] - \chi I[0, 4] \varphi_\gamma[u_0] \Big] \nonumber\\
       &-\frac { f_{3\gamma}\langle g_s^2G^2\rangle } {2^{26} \times 3^6 \times 5 \pi^5}  \Big[  -117 ((4 e_d + 5 e_s) I_1[\mathcal V] + 4 e_u I_2[\mathcal V]) I[0, 4] - 
 4 (16 m_c m_s (e_d - e_u)  I[0, 3] + 9 (e_d + 2 e_s \nonumber\\
 &- e_u) I[0, 4]) \psi^a[
   u_0] \Big]
       \nonumber\\
        &+\frac {e_s\,m_c \langle \bar ss \rangle} {2^{21} \times 3^2 \times 5 \pi^5}\Big[ I_3[\mathcal S] I[0, 5] \Big]\nonumber\\
       &+\frac {  f_{3\gamma}  } {2^{25} \times 3^2 \times 5^2  \times 7 \pi^5}  \Big[ 168 m_c m_s (e_d I_1[\mathcal V] + e_u I_2[\mathcal V]) I[0, 5] + 
 5 ((11 e_d + 36 e_s) I_1[\mathcal V] + 11 e_u I_2[\mathcal V]) I[0, 6] \Big], 
 \end{align}
           \begin{align}
 \rho_3 (\mathrm{M^2},\mathrm{s_0})&=-\frac {183 } {2^{26} \times 5^3 \times 7^2 \pi^7} (e_d + 2 e_s + e_u) I[0,7]\label{F3sonuc} \nonumber\\
        &+\frac { \langle g_s^2G^2\rangle  \langle \bar q q \rangle} {2^{24} \times 3^5 \times 5 \pi^5} (e_d+e_u)\Big[ \Big(-68 m_c \mathbb A[u_0] + 
    3 \big(113 m_c I_3[\mathcal S] + 6 m_c I_3[\mathcal T_2] + 6 m_c I_3[\mathcal T_4] + 
       9 m_c I_4[\mathcal S] 
         \nonumber\\
       &+ 92 m_s I_4[\mathcal S] - 6 m_c I_4[\mathcal T_2] - 
       6 m_c I_4[\mathcal T_4]\big)\Big) I[0, 3]  \Big]\nonumber\\
       &+\frac {e_s \, m_c \langle g_s^2G^2\rangle  \langle \bar ss \rangle} {2^{25} \times 3^4 \times 5 \pi^5}\Big[ (44 \mathbb A[u_0] - 3 I_3[\mathcal S]) I[0, 3] \Big]\nonumber
       \end{align}          
              \begin{align}
       &+\frac { f_{3\gamma}\langle g_s^2G^2\rangle } {2^{30} \times 3^4 \times 5 \pi^5}  \Big[ (-60 (e_d + e_u) I_1[\mathcal A] - 1070( e_d + 2  e_s +  e_u) I_1[
     \mathcal V] + 2 (e_d + e_u) (30 I_2[\mathcal A] + 193 I_2[\mathcal V]) \nonumber\\
       &+ 
   128 ( e_d +2 e_s +  e_u) \psi^a[u_0]) I[0, 4] \Big]\nonumber
       \nonumber\\
      &-\frac { \langle g_s^2G^2\rangle  \langle \bar q q \rangle \chi} {2^{24} \times 3^5 \times 5 \pi^5} (e_d+e_u)\Big[  - 
       6 m_c I_4[\mathcal T_4] I[0, 3] + 
  (56 m_c + 81 m_s) I[0, 4] \varphi_\gamma[u_0] \Big]\nonumber\\
        &-\frac {e_s \, m_c \langle g_s^2G^2\rangle  \langle \bar ss \rangle \chi} {2^{25} \times 3^4 \times 5 \pi^5}\Big[   41  I[0, 4] \varphi_\gamma[u_0] \Big]\nonumber\\
                &+\frac {  \langle \bar q q \rangle} {2^{25} \times 3 \times 5^2 \times 7 \pi^5} (e_d+e_u)\Big[ \Big (306 m_s \mathbb A[u_0] + 175 m_c I_3[\mathcal S] + 
    5 m_s \big (15 I_3[\mathcal S] - 27 I_3[\mathcal T_1] + 27 I_3[\mathcal T_2] \nonumber\\
       &+ 
       127 I_4[\mathcal S] + 27 I_4[\mathcal T_1] - 27 I_4[\mathcal T_2]\big) - 
    21 m_c \big (I_3[\mathcal T_1] - 5 I_3[\mathcal T_2] - 4 I_3[\mathcal T_4] + 5 I_4[\mathcal S] - 
        I_4[\mathcal T_1] + 5 I_4[\mathcal T_2] \nonumber\\
       &+ 4 I_4[\mathcal T_4]\big)\Big) I[0, 5]  \Big]\nonumber\\
        &+\frac {e_s\,m_s \langle \bar ss \rangle} {2^{22} \times 5^2 \pi^5}\Big[ I_3[\mathcal S] I[0, 5] \Big]\nonumber\\
                    &-\frac {  \langle \bar q q \rangle \chi} {2^{25} \times 3 \times 5^2 \times 7 \pi^5} (e_d+e_u)\Big[   
 172  m_s  \varphi_\gamma[u_0] I[0, 6] \Big]\nonumber\\
              &+\frac { f_{3\gamma} } {2^{27} \times 3^2 \times 5^2 \times 7 \pi^5}  \Big[ 45 (e_d + e_u) (14 m_c m_s I[0, 5] + 
    3 I[0, 6]) I_1[\mathcal A] + (-567 m_c m_s (e_d + e_u)  I[0, 5]\nonumber\\
       & + 
    5 (e_d + 2 e_s + e_u) I[0, 6])  I_1[
   \mathcal V] + (e_d + 
    e_u) (-126 m_c m_s (5 I_2[\mathcal A] + 9 I_2[\mathcal V]) I[0, 5] + 
    5 (-27 I_2[\mathcal A] \nonumber\\
       &+ 194 I_2[\mathcal V]) I[0, 6]) - 
 1032 (e_d + 2 e_s + e_u) I[0, 6] \psi^a[u_0] \Big], 
 \end{align}          
              \begin{align}
 \rho_4 (\mathrm{M^2},\mathrm{s_0})&=-\frac {61 e_c } {2^{23}\times 3 \times 5^3 \times 7^2 \pi^7}  I[0,7]\nonumber\\
        &+\frac { \langle g_s^2G^2\rangle  \langle \bar q q \rangle} {2^{25} \times 3^6 \times 5 \pi^5} (e_d+e_u)\Big[ \Big(16 (31 m_c - 11 m_s) \mathbb A[u_0] - 
    9 \big(1093 m_c I_3[\mathcal S] - 21 m_c I_3[\mathcal T_1] + 21 m_c I_3[\mathcal T_2]  \nonumber\\
    &- 39 m_c I_4[\mathcal S] - 184 m_s I_4[\mathcal S] + 
       21 m_c I_4[\mathcal T_1] - 21 m_c I_4[\mathcal T_2]\big)\Big) I[0, 3] \Big]\nonumber\\
         &+\frac {e_s \, m_c \langle g_s^2G^2\rangle  \langle \bar ss \rangle} {2^{24} \times 3^5 \times 5 \pi^5}\Big[(44 \mathbb A[u_0] + 9 I_3[\mathcal S]) I[0, 3] - 32 \chi I[0, 4] \varphi_\gamma[u_0] \Big]\nonumber\\
       &+\frac { f_{3\gamma}\langle g_s^2G^2\rangle } {2^{28} \times 3^6 \times 5 \pi^5}  \Big[ -9 (( e_d + 2 e_s + e_u) I_1[\mathcal V] + 338 (e_d + e_u) I_2[\mathcal V]) I[0, 4] + 
 16 (66 (e_d + e_u) m_c m_s I[0, 3] \nonumber\\
 &+ (5 e_d + 2 e_s + e_u) I[0, 
      4]) \psi^a[u_0] \Big]
       \nonumber\\
        &+\frac { \langle g_s^2G^2\rangle  \langle \bar q q \rangle \chi} {2^{23} \times 3^6 \times 5 \pi^5} (e_d+e_u)\Big[  
  (-64 m_c + 23 m_s) I[0, 4] \varphi_\gamma[u_0] \Big]\nonumber\\
        &+\frac {  \langle \bar q q \rangle} {2^{23} \times 3 \times 5^2 \times 7 \pi^5} (e_d+e_u)\Big[ ((2 m_c + 15 m_s) I_3[\mathcal S] - (6 m_c + 5 m_s) I_4[\mathcal S]) I[0, 5] \Big]\nonumber\\
        &-\frac {e_s\,m_s \langle \bar ss \rangle} {2^{21} \times 5^2 \pi^5}\Big[ I_3[\mathcal S] I[0, 5] \Big]\nonumber\\
       &+\frac { f_{3\gamma} } {2^{25} \times 3^2 \times 5^2 \times 7 \pi^5}  \Big[ 36 (e_d + e_u) m_c m_s (I_1[\mathcal V] + I_2[\mathcal V]) I[0, 5] - 
 5 \Big((e_d + 2 e_s + e_u) I_1[\mathcal V] \nonumber\\
       &+ (e_d + e_u) I_2[\mathcal V]\Big) I[0, 6] \Big].  \label{F4sonuc}
              \end{align} 
Here, $\varphi_\gamma(u)$, $\psi^v(u)$,
$\psi^a(u)$, ${\cal A}(\alpha_i)$ and ${\cal V}(\alpha_i)$, $h_\gamma(u)$, $\mathbb{A}(u)$, ${\cal S}(\alpha_i)$, ${\cal{\tilde S}}(\alpha_i)$, ${\cal T}_1(\alpha_i)$, ${\cal T}_2(\alpha_i)$, ${\cal T}_3(\alpha_i)$ 
and ${\cal T}_4(\alpha_i)$  are the photon DAs. 
The expressions for the functions $ I[n,m] $ and $ I_i[\mathcal{F}] $ are given in the following form:
\begin{align}
 I[n,m]&= \int_{(2m_c +m_s)^2}^{\rm{s_0}} ds ~ e^{-s/\rm{M^2}}~
 s^n\,(s-(2m_c +m_s)^2)^m,\nonumber\\
 I_1[\mathcal{F}]&=\int D_{\alpha_i} \int_0^1 dv~ \mathcal{F}(\alpha_{\bar q},\alpha_q,\alpha_g)
 \delta'(\alpha_ q +\bar v \alpha_g-u_0),\nonumber
 \end{align}          
              \begin{align}
  I_2[\mathcal{F}]&=\int D_{\alpha_i} \int_0^1 dv~ \mathcal{F}(\alpha_{\bar q},\alpha_q,\alpha_g) \delta'(\alpha_{\bar q}+ v \alpha_g-u_0),\nonumber\\
 I_3[\mathcal{F}]&=\int D_{\alpha_i} \int_0^1 dv~ \mathcal{F}(\alpha_{\bar q},\alpha_q,\alpha_g)
 \delta(\alpha_ q +\bar v \alpha_g-u_0),\nonumber\\
   I_4[\mathcal{F}]&=\int D_{\alpha_i} \int_0^1 dv~ \mathcal{F}(\alpha_{\bar q},\alpha_q,\alpha_g) \delta(\alpha_{\bar q}+ v \alpha_g-u_0).
 \end{align}
 Here, $ \mathcal{F} $ represents the corresponding photon DAs.

It is noteworthy that the Borel transformations applied in the expressions above have been carried out in accordance with the following relations:
\begin{align}
 \mathcal{B}\bigg\{ \frac{1}{\big[ [p^2-m^2_i][(p+q)^2-m_f^2] \big]}\bigg\} \rightarrow e^{-m_i^2/\rm{M_1^2}-m_f^2/\rm{M_2^2}}
\end{align}
in the hadronic side, and 
\begin{align}
 \mathcal{B}\bigg\{ \frac{1}{\big(m^2- \bar u p^2-u(p+q)^2\big)^{\alpha}}\bigg\} \rightarrow (\rm{M^2})^{(2-\alpha)} \delta (u-u_0)e^{-m^2/\rm{M^2}},
\end{align}
 in the QCD side, in which we make use of
\begin{align*}
 {\rm{M^2}}= \frac{\rm{M_1^2} \rm{M_2^2}}{\rm{M_1^2}+\rm{M_2^2}}, ~~~~~~
 u_0= \frac{\rm{M_1^2}}{\rm{M_1^2}+\rm{M_2^2}}.
\end{align*}
In this context, $ \rm{M_1^2} $ and $ \rm{M_2^2} $ denote the Borel parameters associated with the initial and final $P_{\psi s}^{\Lambda}$ states, respectively. Since the same $P_{\psi s}^{\Lambda}$ state appears in both the initial and final configurations, we set $ \rm{M_1^2} = \rm{M_2^2} = 2\rm{M^2} $ and $ u_0 = \frac{1}{2} $. This choice ensures that the single dispersion approximation effectively suppresses contributions from higher resonances and the continuum. For further details regarding this procedure, see \cite{Ozdem:2024dbq}.

\section{Photon Distribution Amplitudes}\label{appb}

For completeness, we present in this appendix the explicit forms of the non-perturbative matrix elements that appear on the QCD side of the LCSR calculations, together with the explicit functional forms of the photon DAs entering these matrix elements. The conventions are those of~\cite{Ball:2002ps}, to which we refer the reader for further details.

The matrix elements involving photon DAs are defined as $\langle \gamma(q) | \bar{q}(x) \Gamma_i q(0) | 0 \rangle$ and $\langle \gamma(q) | \bar{q}(x) \Gamma_i G_{\mu\nu} q(0) | 0 \rangle$, where $\Gamma_i$ denotes the relevant Dirac matrix structures. Following the conventions established in~\cite{Ball:2002ps}, these matrix elements take the following forms:
\begin{eqnarray}
\langle \gamma(q) \vert  \bar q(x) \gamma_\mu q(0) \vert 0 \rangle
&=& e_q f_{3 \gamma} \left(\varepsilon_\mu - q_\mu \frac{\varepsilon
x}{q x} \right) \int_0^1 du\, e^{i \bar u q x} \psi^v(u), \\
\langle \gamma(q) \vert \bar q(x) \gamma_\mu \gamma_5 q(0) \vert 0
\rangle  &=& - \frac{1}{4} e_q f_{3 \gamma} \epsilon_{\mu \nu \alpha
\beta } \varepsilon^\nu q^\alpha x^\beta \int_0^1 du\, e^{i \bar u q
x} \psi^a(u), \\
\langle \gamma(q) \vert  \bar q(x) \sigma_{\mu \nu} q(0) \vert  0
\rangle  &=& -i e_q \langle \bar q q \rangle (\varepsilon_\mu q_\nu - \varepsilon_\nu
q_\mu) \int_0^1 du\, e^{i \bar u qx} \left(\chi \varphi_\gamma(u) +
\frac{x^2}{16} \mathbb{A} (u) \right) \nonumber \\ 
&&-\frac{i}{2(qx)}  e_q \langle\bar q q\rangle \left[x_\nu \left(\varepsilon_\mu - q_\mu
\frac{\varepsilon x}{qx}\right) - x_\mu \left(\varepsilon_\nu -
q_\nu \frac{\varepsilon x}{q x}\right) \right] \int_0^1 du\, e^{i \bar
u q x} h_\gamma(u),  \\
\langle \gamma(q) | \bar q(x) g_s G_{\mu \nu} (v x) q(0) \vert 0
\rangle &=& -i e_q \langle \bar q q \rangle \left(\varepsilon_\mu q_\nu - \varepsilon_\nu
q_\mu \right) \int {\cal D}\alpha_i\, e^{i (\alpha_{\bar q} + v
\alpha_g) q x} {\cal S}(\alpha_i),  \\
\langle \gamma(q) | \bar q(x) g_s \tilde G_{\mu \nu}(v
x) i \gamma_5  q(0) \vert 0 \rangle &=& -i e_q \langle \bar q q \rangle \left(\varepsilon_\mu q_\nu -
\varepsilon_\nu q_\mu \right) \int {\cal D}\alpha_i\, e^{i
(\alpha_{\bar q} + v \alpha_g) q x} \tilde {\cal S}(\alpha_i),  \\
\langle \gamma(q) \vert \bar q(x) g_s \tilde G_{\mu \nu}(v x)
\gamma_\alpha \gamma_5 q(0) \vert 0 \rangle &=& e_q f_{3 \gamma}
q_\alpha (\varepsilon_\mu q_\nu - \varepsilon_\nu q_\mu) \int {\cal
D}\alpha_i\, e^{i (\alpha_{\bar q} + v \alpha_g) q x} {\cal
A}(\alpha_i), 
\end{eqnarray}          
             \begin{eqnarray}
\langle \gamma(q) \vert \bar q(x) g_s G_{\mu \nu}(v x) i
\gamma_\alpha q(0) \vert 0 \rangle &=& e_q f_{3 \gamma} q_\alpha
(\varepsilon_\mu q_\nu - \varepsilon_\nu q_\mu) \int {\cal
D}\alpha_i\, e^{i (\alpha_{\bar q} + v \alpha_g) q x} {\cal
V}(\alpha_i), \\
\langle \gamma(q) \vert \bar q(x)
\sigma_{\alpha \beta} g_s G_{\mu \nu}(v x) q(0) \vert 0 \rangle  &=&
e_q \langle \bar q q \rangle \left\{
        \left[\left(\varepsilon_\mu - q_\mu \frac{\varepsilon x}{q x}\right)\left(g_{\alpha \nu} -
        \frac{1}{qx} (q_\alpha x_\nu + q_\nu x_\alpha)\right) \right. \right. q_\beta \nonumber \\ 
         && -
         \left(\varepsilon_\mu - q_\mu \frac{\varepsilon x}{q x}\right)\left(g_{\beta \nu} -
        \frac{1}{qx} (q_\beta x_\nu + q_\nu x_\beta)\right) q_\alpha
\nonumber \\
&&-
         \left(\varepsilon_\nu - q_\nu \frac{\varepsilon x}{q x}\right)\left(g_{\alpha \mu} -
        \frac{1}{qx} (q_\alpha x_\mu + q_\mu x_\alpha)\right) q_\beta \nonumber \\
&&+
         \left. \left(\varepsilon_\nu - q_\nu \frac{\varepsilon x}{q\cdot x}\right)\left( g_{\beta \mu} -
        \frac{1}{qx} (q_\beta x_\mu + q_\mu x_\beta)\right) q_\alpha \right]
   \int {\cal D}\alpha_i\, e^{i (\alpha_{\bar q} + v \alpha_g) qx} {\cal T}_1(\alpha_i) \nonumber \\
 &&+
        \left[\left(\varepsilon_\alpha - q_\alpha \frac{\varepsilon x}{qx}\right)
        \left(g_{\mu \beta} - \frac{1}{qx}(q_\mu x_\beta + q_\beta x_\mu)\right) \right. q_\nu \nonumber \\
        &&
-
         \left(\varepsilon_\alpha - q_\alpha \frac{\varepsilon x}{qx}\right)
        \left(g_{\nu \beta} - \frac{1}{qx}(q_\nu x_\beta + q_\beta x_\nu)\right)  q_\mu \nonumber \\
          && -
         \left(\varepsilon_\beta - q_\beta \frac{\varepsilon x}{qx}\right)
        \left(g_{\mu \alpha} - \frac{1}{qx}(q_\mu x_\alpha + q_\alpha x_\mu)\right) q_\nu \nonumber \\
      &&+
         \left. \left(\varepsilon_\beta - q_\beta \frac{\varepsilon x}{qx}\right)
        \left(g_{\nu \alpha} - \frac{1}{qx}(q_\nu x_\alpha + q_\alpha x_\nu) \right) q_\mu
        \right]      
    \int {\cal D} \alpha_i\, e^{i (\alpha_{\bar q} + v \alpha_g) qx} {\cal T}_2(\alpha_i) \nonumber \\
&&+\frac{1}{qx} (q_\mu x_\nu - q_\nu x_\mu)
        (\varepsilon_\alpha q_\beta - \varepsilon_\beta q_\alpha)
    \int {\cal D} \alpha_i\, e^{i (\alpha_{\bar q} + v \alpha_g) qx} {\cal T}_3(\alpha_i) \nonumber \\ &&+
        \left. \frac{1}{qx} (q_\alpha x_\beta - q_\beta x_\alpha)
        (\varepsilon_\mu q_\nu - \varepsilon_\nu q_\mu)
    \int {\cal D} \alpha_i\, e^{i (\alpha_{\bar q} + v \alpha_g) qx} {\cal T}_4(\alpha_i)
                        \right\}.
\end{eqnarray}

In these expressions, the functions $\varphi_\gamma(u)$, $\psi^v(u)$, $\psi^a(u)$, ${\cal A}(\alpha_i)$, ${\cal V}(\alpha_i)$, $h_\gamma(u)$, $\mathbb{A}(u)$, ${\cal S}(\alpha_i)$, $\tilde {\cal S}(\alpha_i)$, ${\cal T}_1(\alpha_i)$, ${\cal T}_2(\alpha_i)$, ${\cal T}_3(\alpha_i)$, and ${\cal T}_4(\alpha_i)$ represent the photon DAs of different twists. Specifically, $\varphi_\gamma(u)$ corresponds to the leading twist-2 DA, while $\psi^v(u)$, $\psi^a(u)$, ${\cal A}(\alpha_i)$, and ${\cal V}(\alpha_i)$ are associated with twist-3 contributions. The remaining functions $h_\gamma(u)$, $\mathbb{A}(u)$, ${\cal S}(\alpha_i)$, $\tilde {\cal S}(\alpha_i)$, ${\cal T}_1(\alpha_i)$, ${\cal T}_2(\alpha_i)$, ${\cal T}_3(\alpha_i)$, and ${\cal T}_4(\alpha_i)$ encode the twist-4 effects.

The integration measure ${\cal D} \alpha_i$ appearing in the multi-variable distributions is defined as
\begin{eqnarray}
\int {\cal D} \alpha_i = \int_0^1 d \alpha_{\bar q} \int_0^1 d
\alpha_q \int_0^1 d \alpha_g \,\delta(1-\alpha_{\bar
q}-\alpha_q-\alpha_g). 
\end{eqnarray}

The explicit functional forms of the photon DAs employed in our analysis are given by~\cite{Ball:2002ps}:
\begin{eqnarray}
\varphi_\gamma(u) &=& 6 u \bar u \left( 1 + \varphi_2(\mu)
C_2^{\frac{3}{2}}(u - \bar u) \right), \nonumber \\
\psi^v(u) &=& 3 \left(3 (2 u - 1)^2 -1 \right)+\frac{3}{64} \left(15
w^V_\gamma - 5 w^A_\gamma\right)
                        \left(3 - 30 (2 u - 1)^2 + 35 (2 u -1)^4
                        \right), \nonumber \\
\psi^a(u) &=& \left(1- (2 u -1)^2\right)\left(5 (2 u -1)^2 -1\right)
\frac{5}{2}
    \left(1 + \frac{9}{16} w^V_\gamma - \frac{3}{16} w^A_\gamma
    \right), \nonumber \\
h_\gamma(u) &=& - 10 \left(1 + 2 \kappa^+\right) C_2^{\frac{1}{2}}(u
- \bar u), \nonumber \\
\mathbb{A}(u) &=& 40 u^2 \bar u^2 \left(3 \kappa - \kappa^+
+1\right)  +
        8 (\zeta_2^+ - 3 \zeta_2) \left[u \bar u (2 + 13 u \bar u) +  2 u^3 (10 -15 u + 6 u^2) \ln(u)\right. \nonumber \\ &&  \left.
                + 2 \bar u^3 (10 - 15 \bar u + 6 \bar u^2)
        \ln(\bar u) \right], \nonumber \\
{\cal A}(\alpha_i) &=& 360 \alpha_q \alpha_{\bar q} \alpha_g^2
        \left(1 + w^A_\gamma \frac{1}{2} (7 \alpha_g - 3)\right), \nonumber \\
        {\cal V}(\alpha_i) &=& 540\, w^V_\gamma (\alpha_q - \alpha_{\bar q})
\alpha_q \alpha_{\bar q}
                \alpha_g^2, \nonumber \\
{\cal T}_1(\alpha_i) &=& -120 (3 \zeta_2 + \zeta_2^+)(\alpha_{\bar
q} - \alpha_q)
        \alpha_{\bar q} \alpha_q \alpha_g, \nonumber 
        \end{eqnarray}          
             \begin{eqnarray}
{\cal T}_2(\alpha_i) &=& 30 \alpha_g^2 (\alpha_{\bar q} - \alpha_q)
    \left((\kappa - \kappa^+) + (\zeta_1 - \zeta_1^+)(1 - 2\alpha_g) +
    \zeta_2 (3 - 4 \alpha_g)\right), \nonumber \\
{\cal T}_3(\alpha_i) &=& - 120 (3 \zeta_2 - \zeta_2^+)(\alpha_{\bar
q} -\alpha_q)
        \alpha_{\bar q} \alpha_q \alpha_g, \nonumber \\
{\cal T}_4(\alpha_i) &=& 30 \alpha_g^2 (\alpha_{\bar q} - \alpha_q)
    \left((\kappa + \kappa^+) + (\zeta_1 + \zeta_1^+)(1 - 2\alpha_g) +
    \zeta_2 (3 - 4 \alpha_g)\right), \nonumber \\
{\cal S}(\alpha_i) &=& 30\alpha_g^2\left[(\kappa +
\kappa^+)(1-\alpha_g)+(\zeta_1 + \zeta_1^+)(1 - \alpha_g)(1 -
2\alpha_g)+\zeta_2\left(3 (\alpha_{\bar q} - \alpha_q)^2-\alpha_g(1 - \alpha_g)\right)\right], \nonumber \\
\tilde {\cal S}(\alpha_i) &=&-30\alpha_g^2\left[(\kappa -\kappa^+)(1-\alpha_g)+(\zeta_1 - \zeta_1^+)(1 - \alpha_g)(1 -
2\alpha_g)+\zeta_2 \left(3 (\alpha_{\bar q} -\alpha_q)^2-\alpha_g(1 - \alpha_g)\right)\right].
\end{eqnarray}

The numerical values of the parameters entering these DAs are taken as $\varphi_2(1~\mathrm{GeV}) = 0$, $w^V_\gamma = 3.8 \pm 1.8$, $w^A_\gamma = -2.1 \pm 1.0$, $\kappa = 0.2$, $\kappa^+ = 0$, $\zeta_1 = 0.4$, and $\zeta_2 = 0.3$, following the conventions of~\cite{Ball:2002ps}.

\end{widetext}

\bibliographystyle{elsarticle-num}
\bibliography{Pc4338MDM.bib}

\end{document}